\documentclass[10pt,conference]{IEEEtran}
\usepackage{amssymb}
\usepackage{amstext}
\usepackage{amsmath}
\usepackage{epsfig}
\usepackage{epic,eepic}
\input{psfig.sty}

\newtheorem{theorem}{Theorem}[section]
\newtheorem{lemma}[theorem]{Lemma}

\newcommand{\qed}{\nobreak \ifvmode \relax \else
     \ifdim\lastskip<1.5em \hskip-\lastskip
      \hskip1.5em plus0em minus0.5em \fi \nobreak
      \vrule height0.75em width0.5em depth0.25em\fi}

\newcommand{\xS}{\mbox{\boldmath$S$}}  
  
\begin{document}
\title{Duality and Stability Regions of Multi-rate Broadcast and Multiple Access Networks }

\author{\authorblockN{Viveck R. Cadambe and Syed A. Jafar}
\authorblockA{Electrical Engineering and Computer Science\\
University of California, Irvine, CA 92697-2625\\
Email: {vcadambe@uci.edu, syed@uci.edu}}} \maketitle
\IEEEpeerreviewmaketitle

\begin{abstract}

We characterize stability regions of two-user fading Gaussian multiple access (MAC) and broadcast (BC) networks with centralized scheduling. The data to be transmitted to the users is encoded into codewords of fixed length. The rates of the codewords used are restricted to a fixed set of finite cardinality. With successive decoding and interference cancellation at the receivers, we find the set of arrival rates that can be stabilized over the MAC and BC networks. In MAC and BC networks with \textit{average} power constraints, we observe that the duality property that relates the MAC and BC information theoretic capacity regions extend to their stability regions as well. In MAC and BC networks with \textit{peak} power constraints, the union of stability regions of dual MAC networks is found to be strictly contained in the BC stability region. 
\end{abstract}

\section{Introduction}

Traditionally, information theorists and the networking community have taken two different approaches to the study of communication networks. The information theoretic approach explores the \emph{capacity region} \cite{thomas_cover} of a network by optimally designing the encoding and decoding schemes while abstracting the higher layers to infinitely backlogged queues. The networking approach on the other hand explores the \emph{stability region} \cite{gallagher:networks} of a network by optimizing the scheduling and routing algorithms while abstracting the physical layer encoding and decoding schemes to server processes. 

Both networking and information theory perspectives have lead to many important results. For example, the networking approach has found stability-optimal scheduling and routing algorithms \cite{mlwdf,mneely_thesis,mneely:avgp} such as the maximum weight match scheduling and dynamic back-pressure routing algorithms. One of the remarkable successes of information theory, on the other hand, has been the successful characterization of the capacity regions of the Gaussian multiple access (MAC) and broadcast (BC) channels. Information theory has also revealed an elegant duality relationship between the MAC and BC capacity regions \cite{goldsmith_duality}. 

Both networking and information theory perspectives have their own limitations. A common limitation of the networking research has been that sophisticated encoding/decoding techniques introduced by information theory, e.g. successive decoding, are overlooked. The utility of information theoretic results on the other hand is limited by the numerous underlying ideal assumptions such as infinite backlog, continuous rate adaptation and infinitely long codewords. In practice communication systems are designed with a finite set of fixed rate codebooks. The rate adaptation algorithm picks one of the available rates depending on channel and interference conditions, for example \cite{3gpp} \cite{ieee:edge}.

In this work, we combine the strengths of information theory and networking approaches to study wireless broadcast and multiple access networks. Motivated by the networking approach, we study  the \emph{stability regions} of wireless multiple access and broadcast networks when only limited (finite) sets of rates are available to the physical layer. Motivated by the information theoretic approach, we allow the physical layer to use some of the optimal multiuser encoding and decoding schemes for the MAC and BC. In particular we allow independent encoding and successive decoding with successive interference cancellation for both these networks. Note that independent encoding and successive decoding suffice to achieve all points within the information theoretic capacity region of the Gaussian MAC and BC.  Also, motivated by the information theoretic duality of the MAC and BC capacity regions we explore the existence of similar duality relationships for the stability regions of the  multi-rate MAC and BC networks.

 There has been extensive work on the design of scheduling and routing algorithms aware of both the channel state and queue state \cite{cross_survey1}, \cite{cross_survey2}. \cite{mlwdf} proves the modified largest-weighted delay first (MLWDF) scheduling policy for broadcast networks without interference cancellation at the receivers using the concept of Lyapunov drift. \cite{mneely_thesis} extends this to the optimal dynamic back-pressure scheduling and routing algorithm to generic multi-hop networks with peak power constraints. \cite{mneely:avgp} extends these scheduling policies to networks with average power constraints and attempts to find distributed implementations of these policies in networks. The dynamic back pressure algorithm is adapted to cooperative networks in \cite{eyeh:routing_relay}. 
 Associated with the problem of finding optimal scheduling policies is the problem of determining network capacity and stability regions of communication networks. The stability and network capacity regions of general networks with peak power constraints have been found (Refer \cite{mneely:book}).  \cite{ephremides:randommac} and \cite{ltong:aloha} characterize stability regions of multiple access channels with a probabilistic multi-packet reception model (MPR) with distributed scheduling policies - the ALOHA protocol in particular. \cite{ltong:capstab} studies capacity and stability regions of certain large regular wireless networks with the MPR model. Early work in combining information theory with the networking approach can be found in \cite{telatar_qthmac}, \cite{ephremides_infotnetworksunion}. Much like our work, \cite{eyeh_macnetcap} incorporates queues into the information theoretic framework and explores the equivalence of stability regions of a multiple access networks and their ergodic capacity regions; however, unlike our work, it does not assume a finite number of codebooks.  

 The rest of the paper is organized as follows. Section \ref{background} defines and discusses stability and networks capacity regions of networks. Section \ref{sysmodel} presents the system model and the problem is defined in Section \ref{problem_def}. Stability regions of MAC and BC networks with peak power constraints are found in Section \ref{stability:peakp}. Section \ref{macbcavgp} describes techniques on finding stability regions of MAC and BC networks with average power constraints. Section \ref{duality} presents duality results between MAC and BC stability regions. The stability characterization of these multi-rate networks leads to the problem of design of optimal codebooks discussed in Section \ref{optimal_codebooks}. Section \ref{conclusion} concludes with some interesting directions of future work.


\section{Background}
\label{background}

We review two equivalent definitions of the stability region.
\subsection{Stability region of a network}
A stationary stochastic process $Q(t)$ is defined to be \textit{stable} if 
$$ \lim_{x \to \infty} \limsup_{t \to \infty} \mbox{Pr}[Q(t) > x]= 0.$$
A communication network is defined to be stable if the stochastic processes representing its individual queue states are stable. The \textit{stability region} of a network is defined to be the closure of the set of all arrival rate matrices that can be stabilized by some centralized scheduling and routing policy with complete knowledge of the arrival statistics, the queue state and other parameters of the network such as channel conditions, distance between nodes etc. 


\subsection{Equivalent definition based on infinitely backlogged nodes} 
It is shown in \cite{ltong:capstab} that under very mild assumptions that are true for most communication networks (i.e. interference does not increase capacity), the stability region is equivalent to the \emph{network capacity region} (distinct from information theoretic capacity region) defined as follows.

The network layer capacity region is defined as the set of all \emph{average departure rate} matrices that can be achieved in a network assuming that all nodes are \textbf{infinitely backlogged} taking into consideration all possible centralized scheduling routing and resource allocation strategies in the network with possibly complete knowledge of the parameters of the network such as channel conditions, distance between nodes etc.

It must be noted that the information theoretic capacity region is different from the network capacity region; the definition of the former includes information theoretic assumptions such as asymptotically long codewords and uncountable number of codebooks to choose from, whereas the latter definition is more general.  
The equivalence of stability region and network capacity region of a network established in \cite{ltong:capstab} is important because the latter is formulated under the assumption of infinitely backlogged nodes. The infinite backlog model eliminates the dependence on instantaneous queue states from the stability optimal scheduling and routing algorithms. 

\section{The system model}
\label{sysmodel}
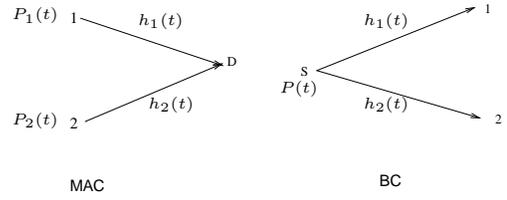
\begin{figure}[!tbp]
\centerline{\setlength{\unitlength}{0.00036667in}
\begingroup\makeatletter\ifx\SetFigFont\undefined%
\gdef\SetFigFont#1#2#3#4#5{%
  \reset@font\fontsize{#1}{#2pt}%
  \fontfamily{#3}\fontseries{#4}\fontshape{#5}%
  \selectfont}%
\fi\endgroup%
{\renewcommand{\dashlinestretch}{30}
\begin{picture}(6992,2700)(0,-10)
\path(4350,1725)(6600,2625)
\path(6499.725,2552.579)(6600.000,2625.000)(6477.441,2608.287)
\path(4350,1725)(6675,1050)
\path(6551.394,1054.647)(6675.000,1050.000)(6568.123,1112.268)
\path(1041,997)(2991,1822)
\path(2892.173,1747.614)(2991.000,1822.000)(2868.795,1802.872)
\path(975,2475)(3000,1800)
\path(2876.671,1809.487)(3000.000,1800.000)(2895.645,1866.408)
\put(3075,1800){\makebox(0,0)[lb]{{\SetFigFont{5}{6.0}{\rmdefault}{\mddefault}{\updefault}D}}}
\put(1800,2325){\makebox(0,0)[lb]{{\SetFigFont{6}{7.2}{\rmdefault}{\mddefault}{\updefault}$h_1(t)$}}}
\put(1950,1125){\makebox(0,0)[lb]{{\SetFigFont{6}{7.2}{\rmdefault}{\mddefault}{\updefault}$h_2(t)$}}}
\put(825,2400){\makebox(0,0)[lb]{{\SetFigFont{6}{7.2}{\familydefault}{\mddefault}{\updefault}1}}}
\put(825,900){\makebox(0,0)[lb]{{\SetFigFont{6}{7.2}{\familydefault}{\mddefault}{\updefault}2}}}
\put(0,900){\makebox(0,0)[lb]{{\SetFigFont{6}{7.2}{\familydefault}{\mddefault}{\updefault}$P_2(t)$}}}
\put(0,2400){\makebox(0,0)[lb]{{\SetFigFont{6}{7.2}{\familydefault}{\mddefault}{\updefault}$P_1(t)$}}}
\put(4125,1650){\makebox(0,0)[lb]{{\SetFigFont{5}{6.0}{\familydefault}{\mddefault}{\updefault}S}}}
\put(6750,2550){\makebox(0,0)[lb]{{\SetFigFont{5}{6.0}{\familydefault}{\mddefault}{\updefault}1}}}
\put(6900,975){\makebox(0,0)[lb]{{\SetFigFont{5}{6.0}{\familydefault}{\mddefault}{\updefault}2}}}
\put(5250,75){\makebox(0,0)[lb]{{\SetFigFont{6}{7.2}{\sfdefault}{\mddefault}{\updefault}BC}}}
\put(3825,1350){\makebox(0,0)[lb]{{\SetFigFont{6}{7.2}{\familydefault}{\mddefault}{\updefault}$P(t)$}}}
\put(5025,2325){\makebox(0,0)[lb]{{\SetFigFont{6}{7.2}{\rmdefault}{\mddefault}{\updefault}$h_1(t)$}}}
\put(5025,1125){\makebox(0,0)[lb]{{\SetFigFont{6}{7.2}{\rmdefault}{\mddefault}{\updefault}$h_2(t)$}}}
\put(825,0){\makebox(0,0)[lb]{{\SetFigFont{6}{7.2}{\sfdefault}{\mddefault}{\updefault}MAC}}}
\end{picture}
}}
\caption{2-user Multiple Access and Broadcast Networks}
\label{figure_mac_bc}
\end{figure}

We shall study the stability regions of multi-rate Gaussian multiple access (MAC) and broadcast networks (BC) (Figure \ref{figure_mac_bc}). We first describe the generic system model that is common to both the models of multiple access and broadcast networks in this work. For ease of exposition, we deal with 2 user MAC and BC networks in this work. However the techniques described and the duality results extend to general N-user MAC and BC networks.

The channel is a Rayleigh fading channel with additive white Gaussian noise at the receiver/s (one receiver in the MAC network and multiple receivers in a BC network).  The noise power at all receivers in all the networks considered are assumed to be unity. The channel fading is represented using the gain vector $\vec{h}(t) = (h_1(t),h_2(t))$, each component of the vector corresponding to a particular link in the network. Note that since stability regions and network capacity regions are to be identical to each other, we use the network capacity model of infinite backlog for our analysis. 
The following are important aspects of the system model.

\subsection{Encoding} The transmitters transmit a single fixed-length codeword per time slot (i.e. the length of one time slot is normalized to the length of the codeword). The codeword is chosen from a fixed finite set of Gaussian codebooks. Let sets $\mathcal{R}_i=\{0,R^i_0,R^i_1\ldots R^i_{M_i-1}\}, i=1,2$ indicate the rates of the codebooks available for transmission corresponding to user $i$ (rate $0$ indicating that the user does not transmit). Therefore, the transmitted rate vector in a particular time slot lies in $\mathcal{R}_1 \times \mathcal{R}_2$.
For example, in a MAC network with both users having a fixed codebook of rate $R_0$, the rate vector that can be transmitted in a particular time slot belongs to $\{(0,0), (R_0,0) (0,R_0) (R_0,R_0)\}$ (indicated by $\times$ symbol in Figure \ref{mac_rate2}). 

\subsection{Slow Fading - Long coherence times}  
The channel $\vec{h}(t)$ is ergodic and remains constant during a time slot. The coherence time of the channel is assumed to be long enough (and therefore a time slot is long enough) so that codewords of sufficient length can be transmitted in a given channel state so as to guarantee information theoretic reliability. The slow fading assumption is reasonable for wireless systems with low mobility.

\subsection{Decoding} 
\label{sysmodel-b}
The decoding strategy at the receivers is restricted to successive decoding and interference cancellation from the received signal. The receiver can decode one complete codeword at a time and cancel the corresponding interference and then proceed to decode the next codeword. This successive decoding procedure is carried on until the receiver decodes his own codeword.

For example, in a two user multiple access network, there are only two different ways of decoding corresponding to two possible decoding orders. The codeword corresponding to user 1 can be decoded first, treating the complete codeword corresponding to user 2 as noise. This is followed by cancellation of interference of user 1 and user 2's codeword is subsequently decoded. The receiver could alternately choose to decode using the reversed order i.e user 2 can be decoded first, followed by interference cancellation and then user 1 can be decoded.  Consider a fixed-codebook MAC network i.e a MAC network of two users with each user having a fixed codebook of rate $R_0$. As indicated earlier, one of the four points indicated by `$\times$' symbol in Figure \ref{mac_rate2} are transmitted. Consider a channel state $\vec{h}$ and a power allocation $(P_1(t), P_2(t))$ such that the information theoretic capacity region of the MAC network at this channel state is represented by pentagon $1$ in Figure \ref{mac_rate2}. Then, a codeword of rate $(r_1,r_2)$ that can be successfully transmitted in this channel state and power allocation is achievable with one of two possible decoding orders.  If rate pair $(r_1, r_2)$ is achievable with user 1 decoded first followed by user 2, it must belong to the rectangle with vertices A, B, E, O. Similarly, if rate pair $(r_1, r_2)$ can be achieved with user 2 decoded first, then it must belong to the rectangle F, C, D, O.  Therefore, in time slot $t$, the rate pairs that can be transmitted reliably over the network with this power allocation lie in the `L-shaped' region (traced in thick lines in Fig. \ref{mac_rate2}). It should be noted that in Figure \ref{mac_rate2}, although $(R_0,R_0)$ lies in the pentagon 1 - the pentagon representing the MAC capacity region, it cannot be achieved with either decoding order and is therefore not achievable at a time slot $t$ with this particular power allocation.

Note that in a BC network, in a particular time slot, all points in the BC capacity region can be achieved with successive interference and decoding. The optimal decoding orders at both receivers depends on the channel state.

\section{Problem Definition}
\label{problem_def}
We aim to study stability regions of multiple access and broadcast networks. The discussion in section \ref{background} implies that we can assume that each node is infinitely backlogged and study the set of achievable departure rates. We categorize the problems we solve into the following:
\begin{enumerate}
\item To find the set of supported rate pairs $(r_1,r_2)$ available for a scheduler at a given channel state.
\item Use the results of part 1 to characterize the stability regions of multiple access and broadcast networks.
\item Explore duality relationships between stability regions of multiple access and broadcast networks.
\end{enumerate}

The rate of transmitted codewords are limited by constraints on the power of the transmitted codewords. Power constraints usually appear in the following forms: 
\begin{itemize}
\item Fixed Power constraints - These are power constraints of the the form $P(t) = \tilde{P}, \forall t$. Analysis of systems with this form of power constraints is useful, particularly when the systems are being studied for a short period of time. However, this model may not be realistic for analysis of certain systems over a long duration.
\item Peak Power constraints - Power constraints of the form $P(t) \leq \tilde{P}, \forall t$ are called peak power constraints and model limits on the instantaneous power of the transmitted signal imposed by the limits of the power source and the amplifiers in the system. 
\item Average Power constraints  - Average power constraints are of the form $E[P(t)] \leq \tilde{P}$ and are used to ensure a certain minimum battery life for the users. 
\end{itemize}

\begin{figure}[!tbp]
\resizebox{3.3 in}{2.2 in}{\setlength{\unitlength}{0.00041667in}
\begingroup\makeatletter\ifx\SetFigFont\undefined%
\gdef\SetFigFont#1#2#3#4#5{%
  \reset@font\fontsize{#1}{#2pt}%
  \fontfamily{#3}\fontseries{#4}\fontshape{#5}%
  \selectfont}%
\fi\endgroup%
{\renewcommand{\dashlinestretch}{30}
\begin{picture}(5346,4215)(0,-10)
\path(825,438)(975,288)
\path(975,438)(825,288)
\path(2025,1638)(2175,1488)
\path(2175,1638)(2025,1488)
\path(2025,438)(2175,288)
\path(2175,438)(2025,288)
\path(825,1638)(975,1488)
\path(975,1638)(825,1488)
\path(900,4188)(900,363)
\drawline(2475,2988)(2475,2988)
\path(825,363)(4800,363)
\thicklines
\path(900,963)(3900,963)(3900,363)
	(900,363)(900,963)
\path(900,3363)(1500,3363)(1500,363)
	(900,363)(900,3363)
\thinlines
\path(2625,363)(2625,1863)(1125,3363)
\dashline{60.000}(900,3363)(1125,3363)(1125,363)
	(900,363)(900,3363)
\dashline{60.000}(900,1863)(2625,1863)(2625,363)
	(900,363)(900,1863)
\path(1500,3363)(3900,963)
\path(1575,3288)(1725,3813)
\path(1720.879,3689.375)(1725.000,3813.000)(1663.188,3705.859)
\path(2175,2313)(2625,3063)
\path(2588.985,2944.666)(2625.000,3063.000)(2537.536,2975.536)
\put(675,63){\makebox(0,0)[lb]{{\SetFigFont{6}{7.2}{\familydefault}{\mddefault}{\updefault}$(0,0)$}}}
\put(600,288){\makebox(0,0)[lb]{{\SetFigFont{6}{7.2}{\familydefault}{\mddefault}{\updefault}O}}}
\put(375,3888){\makebox(0,0)[lb]{{\SetFigFont{7}{8.4}{\familydefault}{\mddefault}{\updefault}$R_2$}}}
\put(4350,63){\makebox(0,0)[lb]{{\SetFigFont{7}{8.4}{\familydefault}{\mddefault}{\updefault}$R_1$}}}
\put(0,1413){\makebox(0,0)[lb]{{\SetFigFont{6}{7.2}{\familydefault}{\mddefault}{\updefault}$(0,R_0)$}}}
\put(4050,888){\makebox(0,0)[lb]{{\SetFigFont{6}{7.2}{\familydefault}{\mddefault}{\updefault}C}}}
\put(3825,138){\makebox(0,0)[lb]{{\SetFigFont{6}{7.2}{\familydefault}{\mddefault}{\updefault}D}}}
\put(750,888){\makebox(0,0)[lb]{{\SetFigFont{6}{7.2}{\familydefault}{\mddefault}{\updefault}F}}}
\put(1800,1263){\makebox(0,0)[lb]{{\SetFigFont{6}{7.2}{\familydefault}{\mddefault}{\updefault}$(R_0,R_0)$}}}
\put(1425,3438){\makebox(0,0)[lb]{{\SetFigFont{6}{7.2}{\familydefault}{\mddefault}{\updefault}B}}}
\put(675,3288){\makebox(0,0)[lb]{{\SetFigFont{6}{7.2}{\familydefault}{\mddefault}{\updefault}A}}}
\put(1650,3963){\makebox(0,0)[lb]{{\SetFigFont{7}{8.4}{\sfdefault}{\mddefault}{\updefault}Pentagon 1: supports $\{(0,0),(R_0,0),(0,R_0)\}$}}}
\put(2100,3663){\makebox(0,0)[lb]{{\SetFigFont{7}{8.4}{\sfdefault}{\mddefault}{\updefault}Does NOT support $(R_0,R_0)$ !}}}
\put(2475,3138){\makebox(0,0)[lb]{{\SetFigFont{7}{8.4}{\sfdefault}{\mddefault}{\updefault}Pentagon 2 }}}
\put(2700,2763){\makebox(0,0)[lb]{{\SetFigFont{7}{8.4}{\sfdefault}{\mddefault}{\updefault}Supports $\{(0,0),(R_0,0),(0,R_0),$}}}
\put(4725,2463){\makebox(0,0)[lb]{{\SetFigFont{7}{8.4}{\sfdefault}{\mddefault}{\updefault}$(R_0,R_0)\}$}}}
\put(1800,63){\makebox(0,0)[lb]{{\SetFigFont{6}{7.2}{\familydefault}{\mddefault}{\updefault}$(R_0,0)$}}}
\put(1425,138){\makebox(0,0)[lb]{{\SetFigFont{6}{7.2}{\familydefault}{\mddefault}{\updefault}E}}}
\end{picture}
}}
\caption{Capacity region and set of supported rates of a MAC at particular channel state (i.e $h_1=h_2=1$) for two different power allocations viz. $(10,10)$ and $(10,4)$ }
\label{mac_rate2}
\end{figure}
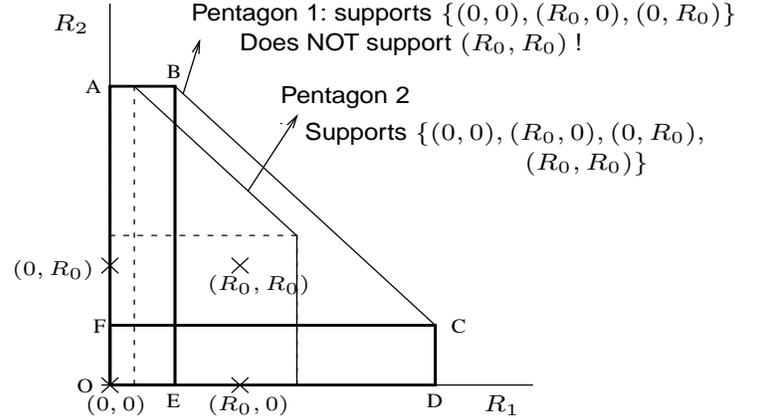

Note that a fixed power constraint is a stronger constraint than a peak power constraint, which is in turn stronger than an average power constraint.  Also note that in our system model, a fixed power constraint is not equivalent to a peak power constraint since better rates can be achieved by transmitting less than the maximum possible power.  For example, in Figure \ref{mac_rate2}, pentagon $2$ represents the MAC capacity region with a power allocation strictly smaller than that which achieves pentagon 1 at the same channel state. We have already argued in Section \ref{sysmodel-b} that $(R_0,R_0)$ cannot be transmitted with the power allocation corresponding to pentagon 1. Since $(R_0,R_0)$ lies in the `L-shaped' region associated with pentagon 2 (plotted with broken lines in Figure \ref{mac_rate2}) it is achievable with the corresponding power allocation.  Therefore reducing the power allocation from that which achieves pentagon $1$ to that corresponding to pentagon $2$ makes $(R_0,R_0)$ achievable in the same channel state.
 In this work, we focus on MAC and BC networks with peak and average power constraints.  
 
 In the remaining part of this work, we use the following notation: $$F(h,\frac{P}{N}) \equiv \log(1+h^2 \frac{P}{N})$$ $$\vec{\chi} = (\chi_1,\chi_2) \equiv (h_1^2,h_2^2)^T = \mbox{power of channel fade} $$ $$\vec{\chi}^{-1} \equiv \mbox{Hadamard reciprocal of } \vec{\chi}$$

\section{Stability Region of MAC and BC Networks with Peak Power constraints}
\label{stability:peakp}
The stability region of a general single hop network with a finite number of channel states and peak power constraint is known (\cite{neely:book}) to be:
\begin{equation}
\label{eqn_stab_peakp}
\Gamma = \displaystyle\sum_{\xS} \mbox{pr}(\xS) \mathrm{Co} ( C(\xS) ) \end{equation}
where $\mbox{pr}(\xS)$ denotes the probability with which the channel takes state $\xS$, $C(\xS)$ denotes the set of all rate vectors that can be transmitted in the network reliably, when the channel state is $\xS$ and $\mathrm{Co}(X)$ denotes the convex hull of set $X$.
For the Gaussian MAC and BC networks, while the Rayleigh fading channel vector $\vec{h}(t) = (h_1(t),h_2(t))$ takes values over a continuum, a finite state space is naturally defined by the partitioning of the channel space based on the rate pairs that can be supported.
Let $\mathcal{C}=2^{\mathcal{R}_1 \times \mathcal {R}_2}$ denote the set of all subsets of $\mathcal{R}_1 \times \mathcal {R}_2$ with cardinality $K=|\mathcal{C}|$. Let $C_i \in \mathcal{C}, i = 1,2 \cdots K$ denote all possible sets of supported rates (or equivalently all possible elements of $2^{\mathcal{R}_1 \times \mathcal{R}_2}$). Then define the state space $S_i, i=1,2 \ldots K $ to be
\begin{eqnarray*}S_i = \{ (h_1,h_2): C_i \in 2^{\mathcal{R}_1 \times \mathcal{R}_2} \mbox{is the \textbf{complete} set of }\\ \mbox{ supported rate vectors, if the channel is }(h_1,h_2) \}\end{eqnarray*}
Note that the set of supported rates is unique to each channel state and equation \ref{eqn_stab_peakp} can now be applied to obtain stability regions.
It must be noted that $S_i$ as defined above may be empty for certain $i$. For example, $S_i = \phi$ if $C_i = \{ (R_m,R_m) \} $ because whenever reliable transmission is possible with rate vector $(R_m, R_m)$, reliable transmission is also possible at rates  $(0,R_m)$, $(R_m,0)$ and $(0,0)$. Therefore $\{(R_m,R_m)\}$ can never be the \textit{complete} set of supported rates for any channel state $(h_1,h_2)$ (in both the MAC and the BC networks).

The definition of $S_i$ can also be understood using Figure \ref{mac_rate2}. Given $(h_1,h_2)$ and powers $P_1 \leq \tilde{P_{1}}, P_2 \leq \tilde{P_{2}}$, an L-shaped region similar to Figure \ref{mac_rate2} can be drawn. Let us represent by  $\Delta(h_1,h_2)$, the union of all such L-shaped regions over $(P_1,P_2)$ pairs satisfying the power constraints. Then $(h_1,h_2) \in S_i$ if and only if the set of all rate vectors from $\mathcal{R}_1 \times \mathcal{R}_2$ that lie in $\Delta(h_1,h_2)$ is precisely $C_i$ i.e $2^{\mathcal{R}_1 \times \mathcal{R}_2} \cap \Delta(h_1,h_2) = C_i$.

\subsection{Characterization of $S_i$ for a MAC network}
\label{macpeakp:rateregion}
A rate vector $(r_1,r_2)$ can be achieved in a MAC network with one of the two decoding orders. If the rate vector is achieved with user 1 decoded before user 2, then conditions \ref{condition:1} and \ref{condition:2} below have to be satisfied.
\begin{equation} 
\label{condition:1}
F(h_2,\frac{\tilde{P_{2}}}{1}) = \log(1+h_2^2 \tilde{P_2} ) \geq r_2
\end{equation}

\begin{equation} 
\label{condition:2}
F(h_1,\frac{\tilde{P_{1}}}{1+h_1^2 P_{2min}}) = \log(1+\frac{h_1 \tilde{P_{1}}^2}{1+h_2^2 P_{2min}} ) \geq r_1
\end{equation}
where $P_{2min}$ is chosen such that $F(h_2,\frac{P_{2min}}{1}) = r_2$.
Condition \ref{condition:2} indicates that the codeword corresponding to user 1 can be decoded when user 2 is transmitting at the minimum possible power(so as to cause minimum interference). Note that this condition can be rewritten as 
$$ F(h_1,\frac{\tilde{P_1}}{2^{r_2}}) \geq r_1$$

Similarly, for achieving $(r_1,r_2)$ with user 2 being decoded first, the channel state $(h_1,h_2)$ must satisfy
$ F(h_1,\frac{\tilde{P_1}}{1}) \geq r_1$ and $ F(h_2,\frac{\tilde{P_2}}{2^{r_1}}) \geq r_2$. 
Define $H_{1\rightarrow2}(S)$ as follows
\begin{eqnarray*} H_{1\rightarrow2}(S) = \{(h_1,h_2) : \mbox{all elements of $S$ can be achieved}\\ \mbox{using decoding order (1,2) \}}\end{eqnarray*} where $S \in 2^{{R_1}\times{R_2}}$.
Using conditions \ref{condition:1} and \ref{condition:2}, it can be shown that \begin{eqnarray*}H_{1\rightarrow2}(S) = \{(h_1,h_2): h_1^2 \geq \frac{(2^{r_1}-1) 2^{r_2}}{\tilde{P_1}} \\ h_2^2 \geq \frac{2^{r_2}-1}{\tilde{P_2}} \forall (r_1,r_2) \in S \}\end{eqnarray*}
Similarly, 
	\begin{eqnarray*}H_{2\rightarrow1}(S) = \{(h_1,h_2) : \mbox{all elements of $S$ can be achieved} \\ \mbox{using decoding order (2,1) }\}\end{eqnarray*}  can be characterized. We can now express $S_i$ as follows.
$$S_i = H_{2\rightarrow1}(C_i) \cup H_{1\rightarrow2}(C_i) \cap \{H_{1\rightarrow2}(C_i^c) \cup H_{2\rightarrow1}(C_i^c)\}^c$$
where $A^{c}$ represents the complement of set $A$. The above equation is merely another form of the definition of $S_i$ suggesting that it is the region of the channel state space where all rate vectors in $C_i$ are supported and the rate vectors outside $C_i$ are not supported. Note that the characterization of $H_{1\rightarrow2}(S)$ indicates that the non-empty channel states $S_i$ can be represented as rectangular regions in the $(h_1,h_2)$ plane (see, for example, Figure \ref{h:peakppart})
The stability region of the MAC network can then be calculated using equation \ref{eqn_stab_peakp}.

\textit{Example : Fixed codebook MAC network}\\
Consider a MAC network with a single fixed codebook of rate $R_0$ at both users, i.e $\mathcal{R} = \mathcal{R}_1 = \mathcal{R}_2 = \{0,R_0\}$. Therefore we have $2^4=16$ possible subsets of $\mathcal{R} \times \mathcal{R}$. However, it is easy to see that, except for five of these subsets, the channel partitions corresponding to all other subsets are empty. These five subsets, represented as $C_i, i=1,2,3,4,5$ and their corresponding channel partitions $S_i$ are described below.
\begin{enumerate}
\item $S_1$ :  This represents the state where the channel gains are so low that transmission is not possible on either channel. i.e $F(h_i, \tilde{P_i}) < R_0, i=1,2$ with $C_1 = \{(0,0)\}$
\item $S_2$ : Transmission is possible only on channel 1 i.e $C_2 = \{(R_0,0), (0,0)\}$ and therefore $F(h_1, \tilde{P_1}) \geq R_0 $, $F(h_2, \tilde{P_2}) < R_0 $
\item $S_3$ : Transmission on channel 2 $C_3 = \{(0,R_0), (0,0)\}$ and $F(h_1, \tilde{P_1}) < R_0 $, $F(h_2, \tilde{P_2}) \geq R_0 $
\item $S_4$ : Transmission is possible individually on both channels AND simultaneously on both channels i.e $C_4 = \mathcal{R} \times \mathcal{R} = \{(0,0), (R_0,0), (0,R_0) (R_0,R_0)\}$.
For $(0,R_0)$ and $(R_0,0)$, we require
$$F(h_i, \tilde{P_i}) \geq R_0, i=1,2 $$
$(R_0,R_0)$ can be achieved with two possible decoding orders. If user 1 is decoded first, user 2 transmits at the minimum power required to decode a codeword of rate $R_0$ so that the user causes the minimum possible interference. Therefore, the condition for successful decoding of $(R_0,R_0)$ with user 1 decoded first can be written as
$$ F(h_1, \frac{\tilde{P_1}}{1+h_2^2 P_{2min}}) \geq R_0 $$
Similarly, if user 2 is decoded first, the condition for successful decoding is.
$ F(h_2,\frac{\tilde{P_2}}{1+h_1^2 P_{1min}}) \geq R_0 $ where $F(h_1, P_{1min}) = R_0$ and $F(h_2, P_{2min}) = R_0$
Therefore, the condition for $\vec{h}$ to lie in $S_4$ can be described by the following equations 
$$F(h_i, \tilde{P_i}) \geq R_0, i=1,2 $$ 
\begin{center}$ F(h_1, \frac{\tilde{P_1}}{2^{R_0})}) \geq R_0 $ or $ F(h_2, \frac{\tilde{P_2}}{2^{R_0})}) \geq R_0 $\end{center}
\item $S_5$ : Transmission is possible individually in both channels, but simultaneous transmission is not possible. i.e
$C_5 = \{(0,0), (R_0,0), (0,R_0)\}$ and
$$F(h_i, \tilde{P_i}) \geq R_0, i=1,2 $$
$$ F(h_1, \frac{\tilde{P_1}}{2^{R_0})}) < R_0 $$
$$ F(h_2, \frac{\tilde{P_2}}{2^{R_0})}) < R_0 $$
\end{enumerate}
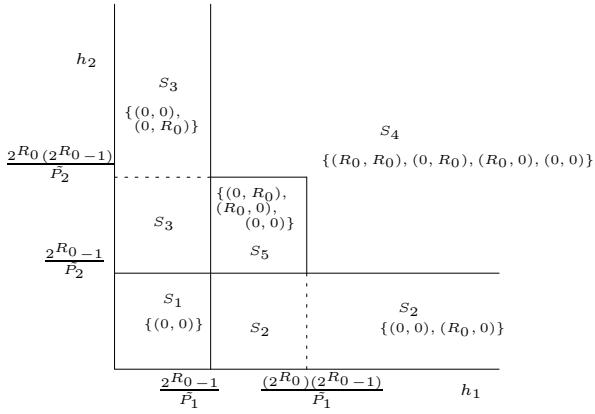
\begin{figure}[!tbp]
{\setlength{\unitlength}{0.00033333in}
\begingroup\makeatletter\ifx\SetFigFont\undefined%
\gdef\SetFigFont#1#2#3#4#5{%
  \reset@font\fontsize{#1}{#2pt}%
  \fontfamily{#3}\fontseries{#4}\fontshape{#5}%
  \selectfont}%
\fi\endgroup%
{\renewcommand{\dashlinestretch}{30}
\begin{picture}(7737,6385)(0,-10)
\path(1725,6358)(1725,658)(6825,658)(7725,658)
\path(3225,658)(3225,6358)
\path(3225,3658)(4725,3658)
\path(4725,3658)(4725,2158)
\dashline{60.000}(1725,3658)(3225,3658)
\dashline{60.000}(4725,2158)(4725,658)
\path(1725,2158)(7725,2158)
\put(5850,4258){\makebox(0,0)[lb]{{\SetFigFont{5}{6.0}{\familydefault}{\mddefault}{\updefault}$S_4$}}}
\put(5850,1183){\makebox(0,0)[lb]{{\SetFigFont{5}{6.0}{\rmdefault}{\mddefault}{\updefault}$\{(0,0),(R_0,0)\}$}}}
\put(2175,1258){\makebox(0,0)[lb]{{\SetFigFont{5}{6.0}{\rmdefault}{\mddefault}{\updefault}$\{(0,0)\}$}}}
\put(600,2083){\makebox(0,0)[lb]{{\SetFigFont{5}{6.0}{\familydefault}{\mddefault}{\updefault}$\frac{2^{R_0}-1}{\tilde{P_2}}$}}}
\put(7125,208){\makebox(0,0)[lb]{{\SetFigFont{6}{7.2}{\familydefault}{\mddefault}{\updefault}$h_1$}}}
\put(2475,1633){\makebox(0,0)[lb]{{\SetFigFont{5}{6.0}{\familydefault}{\mddefault}{\updefault}$S_1$}}}
\put(4950,3808){\makebox(0,0)[lb]{{\SetFigFont{5}{6.0}{\familydefault}{\mddefault}{\updefault}$\{(R_0,R_0),(0,R_0),(R_0,0),(0,0)\}$}}}
\put(3825,1183){\makebox(0,0)[lb]{{\SetFigFont{5}{6.0}{\rmdefault}{\mddefault}{\updefault}$S_2$}}}
\put(6150,1483){\makebox(0,0)[lb]{{\SetFigFont{5}{6.0}{\rmdefault}{\mddefault}{\updefault}$S_2$}}}
\put(3825,2383){\makebox(0,0)[lb]{{\SetFigFont{5}{6.0}{\rmdefault}{\mddefault}{\updefault}$S_5$}}}
\put(2400,5008){\makebox(0,0)[lb]{{\SetFigFont{5}{6.0}{\familydefault}{\mddefault}{\updefault}$S_3$}}}
\put(2325,2833){\makebox(0,0)[lb]{{\SetFigFont{5}{6.0}{\familydefault}{\mddefault}{\updefault}$S_3$}}}
\put(1125,5383){\makebox(0,0)[lb]{{\SetFigFont{6}{7.2}{\familydefault}{\mddefault}{\updefault}$h_2$}}}
\put(0,3583){\makebox(0,0)[lb]{{\SetFigFont{5}{6.0}{\familydefault}{\mddefault}{\updefault}$\frac{2^{R_0}(2^{R_0}-1)}{\tilde{P_2}}$}}}
\put(3300,3283){\makebox(0,0)[lb]{{\SetFigFont{5}{6.0}{\familydefault}{\mddefault}{\updefault}$\{(0,R_0),$}}}
\put(3300,3058){\makebox(0,0)[lb]{{\SetFigFont{5}{6.0}{\familydefault}{\mddefault}{\updefault}$(R_0,0),$}}}
\put(3750,2833){\makebox(0,0)[lb]{{\SetFigFont{5}{6.0}{\familydefault}{\mddefault}{\updefault}$(0,0)\}$}}}
\put(2025,4333){\makebox(0,0)[lb]{{\SetFigFont{5}{6.0}{\familydefault}{\mddefault}{\updefault}$(0,R_0)\}$}}}
\put(2400,58){\makebox(0,0)[lb]{{\SetFigFont{5}{6.0}{\familydefault}{\mddefault}{\updefault}$\frac{2^{R_0}-1}{\tilde{P_1}}$}}}
\put(3975,58){\makebox(0,0)[lb]{{\SetFigFont{5}{6.0}{\familydefault}{\mddefault}{\updefault}$\frac{(2^{R_0})(2^{R_0}-1)}{\tilde{P_1}}$}}}
\put(1875,4558){\makebox(0,0)[lb]{{\SetFigFont{5}{6.0}{\familydefault}{\mddefault}{\updefault}$\{(0,0),$}}}
\end{picture}
}}
\caption{Optimal mapping of channel state $(h_1,h_2)$ to rate sets $C_i$ in the MAC network with peak power constraints}
\label{h:peakppart}
\end{figure}

Figure \ref{h:peakppart} shows the partitioning $S_i$ of the $(h_1,h_2)$ space and the sets of supported rates $C_i$ corresponding to those partitions.

The stability region of this network can be found analytically from the definition in (\ref{eqn_stab_peakp}) and is plotted in Fig. \ref{resultsfig:peakp} for $R_0=1$ and power constraints $\tilde{P_1}$ and $\tilde{P_2}$ satisfying $\tilde{P_1}+\tilde{P_2}=2$. (The variance of the channel fade is assumed to be unity). 
It can be shown that as long as the probabilities $\mbox{pr}(S_i)$  are greater than zero, the stability region is a pentagon. 

\subsection{Characterization of $S_i$ for a BC network}
\label{bcstab:peakp}
\begin{figure}[!tbp]
{\setlength{\unitlength}{0.00033333in}
\begingroup\makeatletter\ifx\SetFigFont\undefined%
\gdef\SetFigFont#1#2#3#4#5{%
  \reset@font\fontsize{#1}{#2pt}%
  \fontfamily{#3}\fontseries{#4}\fontshape{#5}%
  \selectfont}%
\fi\endgroup%
{\renewcommand{\dashlinestretch}{30}
\begin{picture}(6687,6539)(0,-10)
\path(675,6363)(675,513)(6600,513)
\path(675,2013)(6675,2013)
\path(2175,513)(2175,6438)
\dashline{60.000}(2175,2013)(3225,3063)
\path(4484,5577)(2984,4977)
\path(3084.275,5049.421)(2984.000,4977.000)(3106.559,4993.713)
\path(5925,3363)(5625,2688)
\path(5646.322,2809.842)(5625.000,2688.000)(5701.151,2785.473)
\path(3225,3063)(3227,3063)(3232,3062)
	(3241,3060)(3254,3057)(3273,3053)
	(3297,3049)(3326,3043)(3359,3037)
	(3395,3029)(3434,3022)(3474,3014)
	(3515,3006)(3556,2998)(3596,2990)
	(3635,2982)(3672,2975)(3708,2968)
	(3742,2961)(3774,2955)(3805,2949)
	(3835,2944)(3864,2938)(3892,2933)
	(3919,2928)(3946,2923)(3973,2918)
	(4000,2913)(4025,2908)(4051,2904)
	(4077,2899)(4103,2894)(4130,2890)
	(4158,2885)(4186,2880)(4215,2875)
	(4245,2870)(4276,2865)(4307,2860)
	(4338,2855)(4370,2850)(4403,2845)
	(4435,2840)(4468,2835)(4502,2830)
	(4535,2825)(4568,2820)(4602,2815)
	(4635,2811)(4668,2806)(4701,2802)
	(4733,2798)(4765,2794)(4797,2790)
	(4829,2786)(4861,2783)(4893,2779)
	(4925,2775)(4955,2772)(4986,2769)
	(5017,2766)(5049,2763)(5081,2760)
	(5113,2757)(5147,2754)(5181,2751)
	(5216,2748)(5251,2745)(5286,2742)
	(5322,2740)(5359,2737)(5395,2734)
	(5432,2732)(5468,2729)(5504,2726)
	(5541,2724)(5576,2722)(5611,2720)
	(5646,2717)(5680,2715)(5713,2713)
	(5746,2712)(5777,2710)(5808,2708)
	(5838,2707)(5867,2705)(5895,2704)
	(5922,2703)(5949,2702)(5975,2700)
	(6006,2699)(6037,2698)(6067,2697)
	(6097,2696)(6127,2695)(6158,2694)
	(6189,2694)(6220,2693)(6253,2692)
	(6287,2692)(6323,2691)(6359,2691)
	(6396,2690)(6434,2690)(6472,2689)
	(6508,2689)(6543,2689)(6575,2689)
	(6603,2688)(6627,2688)(6645,2688)
	(6659,2688)(6668,2688)(6673,2688)(6675,2688)
\path(3258,3062)(3258,3064)(3257,3069)
	(3255,3078)(3252,3091)(3248,3110)
	(3244,3134)(3238,3163)(3232,3196)
	(3224,3232)(3217,3271)(3209,3311)
	(3201,3352)(3193,3393)(3185,3433)
	(3177,3472)(3170,3509)(3163,3545)
	(3156,3579)(3150,3611)(3144,3642)
	(3139,3672)(3133,3701)(3128,3729)
	(3123,3756)(3118,3783)(3113,3810)
	(3108,3837)(3103,3862)(3099,3888)
	(3094,3914)(3089,3940)(3085,3967)
	(3080,3995)(3075,4023)(3070,4052)
	(3065,4082)(3060,4113)(3055,4144)
	(3050,4175)(3045,4207)(3040,4240)
	(3035,4272)(3030,4305)(3025,4339)
	(3020,4372)(3015,4405)(3010,4439)
	(3006,4472)(3001,4505)(2997,4538)
	(2993,4570)(2989,4602)(2985,4634)
	(2981,4666)(2978,4698)(2974,4730)
	(2971,4762)(2967,4792)(2964,4823)
	(2961,4854)(2958,4886)(2955,4918)
	(2952,4950)(2949,4984)(2946,5018)
	(2943,5053)(2940,5088)(2937,5123)
	(2935,5159)(2932,5196)(2929,5232)
	(2927,5269)(2924,5305)(2921,5341)
	(2919,5378)(2917,5413)(2915,5448)
	(2912,5483)(2910,5517)(2908,5550)
	(2907,5583)(2905,5614)(2903,5645)
	(2902,5675)(2900,5704)(2899,5732)
	(2898,5759)(2897,5786)(2896,5812)
	(2894,5843)(2893,5874)(2892,5904)
	(2891,5934)(2890,5964)(2889,5995)
	(2889,6026)(2888,6057)(2887,6090)
	(2887,6124)(2886,6160)(2886,6196)
	(2885,6233)(2885,6271)(2884,6309)
	(2884,6345)(2884,6380)(2884,6412)
	(2883,6440)(2883,6464)(2883,6482)
	(2883,6496)(2883,6505)(2883,6510)(2883,6512)
\put(75,5013){\makebox(0,0)[lb]{{\SetFigFont{6}{7.2}{\sfdefault}{\mddefault}{\updefault}$h_2$}}}
\put(1350,1413){\makebox(0,0)[lb]{{\SetFigFont{6}{7.2}{\sfdefault}{\mddefault}{\updefault}$S_1$}}}
\put(3750,1338){\makebox(0,0)[lb]{{\SetFigFont{6}{7.2}{\sfdefault}{\mddefault}{\updefault}$S_2$}}}
\put(3375,963){\makebox(0,0)[lb]{{\SetFigFont{5}{6.0}{\familydefault}{\mddefault}{\updefault}${(R_0,0),(0,0)}$}}}
\put(1125,3738){\makebox(0,0)[lb]{{\SetFigFont{6}{7.2}{\sfdefault}{\mddefault}{\updefault}$S_3$}}}
\put(2550,3738){\makebox(0,0)[lb]{{\SetFigFont{6}{7.2}{\sfdefault}{\mddefault}{\updefault}$S_5$}}}
\put(5850,138){\makebox(0,0)[lb]{{\SetFigFont{6}{7.2}{\sfdefault}{\mddefault}{\updefault}$h_1$}}}
\put(1125,963){\makebox(0,0)[lb]{{\SetFigFont{5}{6.0}{\familydefault}{\mddefault}{\updefault}$\{(0,0)\}$}}}
\put(825,4638){\makebox(0,0)[lb]{{\SetFigFont{5}{6.0}{\familydefault}{\mddefault}{\updefault}$\{(0,0),$}}}
\put(3150,5688){\makebox(0,0)[lb]{{\SetFigFont{6}{7.2}{\familydefault}{\mddefault}{\updefault}$\frac{(k-1)}{h_2^2} + \frac{k(k-1)}{h_1^2} = \tilde{P}$}}}
\put(4650,4713){\makebox(0,0)[lb]{{\SetFigFont{5}{6.0}{\familydefault}{\mddefault}{\updefault}$\{(R_0,R_0)(R_0,0),(0,R_0),(0,0)\}$}}}
\put(5625,4338){\makebox(0,0)[lb]{{\SetFigFont{6}{7.2}{\sfdefault}{\mddefault}{\updefault}$S_4$}}}
\put(4800,3438){\makebox(0,0)[lb]{{\SetFigFont{5}{6.0}{\familydefault}{\mddefault}{\updefault}$\frac{(k-1)}{h_1^2} + \frac{k(k-1)}{h_2^2} = \tilde{P}$}}}
\put(3600,2538){\makebox(0,0)[lb]{{\SetFigFont{6}{7.2}{\sfdefault}{\mddefault}{\updefault}$S_5$}}}
\put(900,4263){\makebox(0,0)[lb]{{\SetFigFont{5}{6.0}{\familydefault}{\mddefault}{\updefault}$(0,R_0)\}$}}}
\put(3075,2238){\makebox(0,0)[lb]{{\SetFigFont{5}{6.0}{\familydefault}{\mddefault}{\updefault}$\{(R_0,0),(0,R_0),(0,0)\}$}}}
\put(1950,63){\makebox(0,0)[lb]{{\SetFigFont{6}{7.2}{\familydefault}{\mddefault}{\updefault}$\frac{k-1}{\tilde{P}}$}}}
\put(0,1938){\makebox(0,0)[lb]{{\SetFigFont{6}{7.2}{\familydefault}{\mddefault}{\updefault}$\frac{k-1}{\tilde{P}}$}}}
\end{picture}
}}
\caption{Optimal mapping of channel state $(h_1,h_2)$ to rate sets $C_i$ in the BC network with peak power constraints}
\label{h:peakppartbc}
\end{figure}
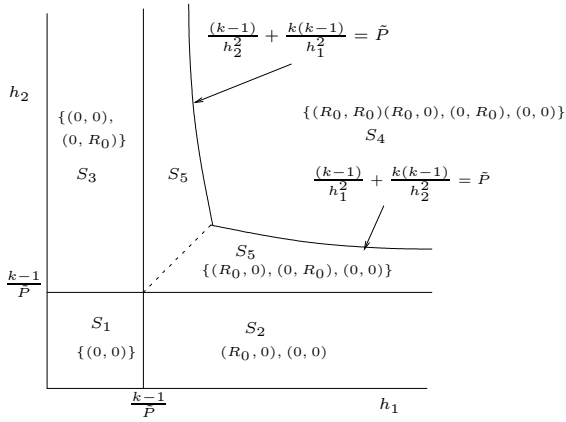

In a BC network, the channel state imposes a decoding order i.e if $h_1 > h_2$, then both users decode the codeword corresponding to user 2 first; node 1 then decodes the codeword corresponding to user 1 after canceling the interference from the codeword corresponding to user 2. Therefore, for $h_1 > h_2$, the condition for achieving codeword $(r_1,r_2)$ can be expressed as
$$ F(h_1, \tilde{P}) \geq r_1$$
$$ F(h_2, \frac{\tilde{P}-P_{1min}}{1+h_2 P_{1min}}) \geq r_2$$
where $P_{1min}$ is chosen such that $F(h_2,P_{1min}) = r_1$.
Define $H_{1,2}(S)$ and $H_{2,1}(S)$ as follows 

\begin{eqnarray*}H_{1,2}(S) = \{(h_1,h_2) : h_1 > h_2, \mbox{all rate vectors of set $S$ can}\\ \mbox{ be achieved in the broadcast channel }\}\end{eqnarray*} 
\begin{eqnarray*}H_{2,1}(S) = \{(h_1,h_2) : h_1 \leq h_2, \mbox{all rate vectors of set $S$ can}\\ \mbox{ be achieved in the broadcast channel }\}\end{eqnarray*} 
$H_{1,2}(S)$ can be characterized as $$H_{1,2}(S) = \{(h_1,h_2): h_1 > h_2, A_{r_1,r_2} \vec{\chi}^{-1} \leq (\tilde{P},\tilde{P})^T, \forall (r_1,r_2) \in S\}$$ where 
$$A_{r_1,r_2} = \left( \begin{array}{cc} 2^{r_1}-1 & 0 \\
		(2^{r_1}-1) 2^{r_2} & (2^{r_2} -1) \end{array} \right) $$
		
and as defined earlier, $\vec{\chi}$ is the power of the channel fade.  $H_{2,1}(S)$ can be similarly characterized. Therefore, 
$$S_i = H_{2,1}(C_i) \cup H_{1,2}(C_i) \cap \{H_{1,2}(C_i^c) \cup H_{1,2}(C_i^c)\}$$
Note that the characterization of $H_{1,2}(S)$ indicates that the boundaries between different $S_i$ in the $(h_1,h_2)$ plane can be expressed as linear equations in $\frac{1}{h_i^2}$ (for example see Figure \ref{h:peakppartbc}).

\textit{Example : Fixed codebook broadcast network}

Consider a two user broadcast network with  fixed codebooks of rate $R_0$ and the transmit power governed by $P(t) \leq \tilde{P}$.  Similar to the MAC case, the channel state is divided into 5 non-empty bins listed below.

In the description below $P_{1min}$ and $P_{2min}$ take values such that  $\log(1+h_1^2 P_{1min}) = R_0$ and $\log(1+h_2^2 P_{2min}) = R_0$
\begin{enumerate}
\item  $S_1$ : $C_1 = \{(0,0)\}$, $\Rightarrow F(h_i, \tilde{P}) < R_0, i=1,2$
\item $S_2$ : $C_2 = \{(0,0),(0,R_0)\}$, $F(h_1, \tilde{P}) < R_0 $
and $F(h_2, \tilde{P}) \geq R_0 $
\item $S_3$ : $C_3 = \{(0,0),(R_0,0)\}$, $\Rightarrow F(h_2, \tilde{P}) < R_0 $ and $F(h_1, \tilde{P}) \geq R_0 $

\item $S_4$ : $ C_4 = \{(0,0) (R_0,0) (0,R_0) (R_0,R_0)\}$
$$F(h_i, \tilde{P}) \geq R_0, i=1,2$$
$$F(h_1, \frac{\tilde{P}-P_{2min}}{1+h_1^2 P_{2min}}) \geq R_0 \mbox{ if }h_1 \leq h_2$$ and
$$F(h_2,\frac{\tilde{P}-P_{1min}}{1+h_2^2 P_{1min}}) \geq R_0 \mbox{ if }h_1 > h_2$$
the latter two equations implying that the weaker user is able to decode the codeword, with the stronger user transmitting at the minimum possible power, so as to cause minimum interference.

\item $S_5$: $ C_5 = \{(0,0) (R_0,0) (0,R_0)\}$\\
$F(h_i, \tilde{P}) \geq R_0, i=1,2$, 
$F(h_1, \frac{\tilde{P}-P_{2min}}{1+h_1^2 P_{2min}}) < R_0 $ and $F(h_2, \frac{\tilde{P}-P_{1min}}{1+h_2^2 P_{1min}}) < R_0 $
\end{enumerate}

Figure \ref{h:peakppartbc} shows the partitioning $S_i$ of the $(h_1,h_2)$ space and corresponding rate sets $C_i$ in the BC network. 
Figure \ref{resultsfig:peakp} has a plot of the stability region of a broadcast network (which is also a pentagon) with $\tilde{P}=2$ and $R_0=1$.

\section{Stability regions of MAC and BC Networks with Average Power Constraints}
\label{macbcavgp}

The problem in finding stability regions of networks with average power constraints is a little more involved because of the following reason. In networks with average power constraints, at any given instant of time, all the available rates can be transmitted by pumping sufficient power. Therefore, the problem also involves finding the optimal power allocation for each channel state, and therefore the simple expression in equation \ref{eqn_stab_peakp} cannot be applied. We therefore use techniques of convex optimization to find the stability region (\cite{syed_cdmamultirate} uses the technique in a similar context).

The stability region is equivalent to the set of all average departure rates assuming infinite backlog. A scheduler can therefore use time-division to stabilize any convex combination of rates in the stability region implying that the stability region is convex. Since the boundary of a convex region can completely be described by its tangents, the boundary of the stability region can be found by finding all tangents of the region. This is equivalent to maximizing $ <\vec{w},\vec{R}>, \forall 0 < w \leq 1$, under the given power constraint, where  $\vec{R} = (\bar{R_1},\bar{R_2})$ represents the average rate vector in the stability region, $\vec{w} = (w,1-w)$ represents the slope of the tangent,  and $<\vec{x},\vec{y}>$ represents the dot product of vectors $x$ and $y$. Note that this formulation is due to the convex nature of the stability region and is independent of the form of the power constraint. In general, the optimal rate allocation strategy may involve transmitting different rate vectors (possibly randomly with different probabilities) at a particular channel state. For example, for fixed codebook MAC networks with peak power constraints, it was optimal to multiplex between $(R_0,0)$ and $(0,R_0)$ when the channel state was in $S_5$. However in networks with average power constraints, if the channel comes from an infinite state space, the optimal strategy transmits a single unique rate vector at a given channel state. We state this formally below.

\begin{lemma}
In MAC and BC networks with average power constraints, if the components of $\vec{h}$ take values from a continuous state space, the optimal rate allocation strategy that maximizes $<\vec{w},\vec{R}>$ associates with a channel state $\vec{h}$ a \emph{unique} rate vector $R(\vec{h}) \in \mathcal{R}_1 \times \mathcal{R}_2$. And therefore, the optimization problem may be expressed as
$$ \max_{\vec{R}(\vec{h}) \in \mathcal{R}_1 \times \mathcal{R}_2} \int_{\vec{h}} <\vec{w},\vec{R}(\vec{h})>  f(\vec{h}) d \vec{h} $$
$$ \mbox{s.t} \int_{\vec{h}} P_i(\vec{h}) f(\vec{h}) d \vec{h}  \leq \tilde{{P}_i}, i\in\{\mbox{set of transmitting nodes}\}$$ 
and $\vec{R}(\vec{h})$ can be reliably transmitted at channel state $\vec{h}$ with powers $P_i(\vec{h})$
\end{lemma}
\begin{proof}
Proof is presented in Appendix \ref{proof:avgp_lemma}.
\end{proof}
The problem reduces to one of finding the optimal $\vec{R}(\vec{h})$ for a given $w$, so that the power constraint is satisfied.
We now treat the MAC and BC networks separately. 

\subsection{MAC Network}
Consider a MAC network with the power constraints at the two nodes described by $E[P_i(t)] \leq \tilde{P_i}, i=1,2$.In the MAC network, for a given channel state, rate vector can be achieved with different transmit powers based on the decoding order. 
Let us denote two possible decoding orders of a two-user MAC network by by $\Pi = \{\pi_1, \pi_2\}$. $\pi_1$ represents the order of decoding user 1 first and then user 2, and $\pi_2$, vice-versa. Let $P_i^{\pi_j}(\vec{r},\vec{h}), i=1,2$ denote the transmit powers of the users required to transmit rate vector $\vec{r}$ at channel state $\vec{h}$ when the decoding order is fixed at $\pi_{j}$. The solution to the optimal rate and power allocation problem is given by


\begin{eqnarray*}
(\vec{R}^{*}(\vec{\chi}),\pi^{*}(\vec{\chi})) = \arg\max_{\vec{r} \in \mathcal{R}_1 \times \mathcal{R}_2,\pi \in \Pi} \{<\vec{w},\vec{r}> \\- \kappa_1 P_1^{\pi}(\vec{r},\vec{\chi}) - \kappa_2 P_2^{\pi}(\vec{r},\vec{\chi}) \}
\end{eqnarray*}
with the Lagrangian multipliers $\kappa_1$ and $\kappa_2$ chosen so that the power constraints of the system are met.
It can be shown (using equations similar to \ref{condition:1} and \ref{condition:2} in \ref{macpeakp:rateregion}) that  
$$P_1^{\pi_1}(\vec{r},\vec{\chi}) = \frac{ 2^{r_2}(2^{r_1}-1)}{h_1^2}$$ $$P_2^{\pi_1}(\vec{r},\vec{h}) = \frac{(2^{r_2}-1)}{h_2^2}$$ where $\vec{r}=(r_1,r_2)$

$$P_1^{\pi_2}(\vec{r},\vec{\chi}) = \frac{ (2^{r_1}-1)}{h_1^2}$$ and $$P_2^{\pi_2}(\vec{r},\vec{h}) = \frac{2^{r_1}(2^{r_2}-1)}{h_2^2}$$ where $\vec{r}=(r_1,r_2)$
The set of rate vectors $(r_1,r_2)\in 2^{\mathcal{R}_1\times \mathcal{R}_2}$ divides the channel state space $\vec{\chi}$ into different regions $S^{\pi}_{\vec{r_1^{*},r_2^{*}}} = \{\vec{\chi} : \vec{R}^{*}(\vec{\chi}) = (r_1^*,r_2^{*}), \pi^{*}(\vec{\chi})\}$. It can be shown that 
$$S^{\pi}_{(r_1^{*},r_2^{*})} = \{\vec{\chi} : A^{\pi}_{r_1,r_2} \vec{\chi}^{-1} \leq (v,v)^T, \forall (r_1,r_2) \in 2^{\mathcal{R}_1 \times \mathcal{R}_2}\}$$
where $v=<\vec{w},(r_1^*-r_1,r_2^*-r_2)>$ and

$$A_{(r_1,r_2)}^{\pi_1} = \left( \begin{array}{ll}
        f_1^1(\vec{r^*}) - f_1^1(\vec{r}) & f_2^1(\vec{r^*}) - f_2^1(\vec{r})\\
        f_1^1(\vec{r^*}) - f_1^2(\vec{r}) & f_2^1(\vec{r^*}) - f_2^2(\vec{r}) \end{array} \right)$$

$$A_{(r_1,r_2)}^{\pi_2} = \left( \begin{array}{ll}
        f_1^2(\vec{r^*}) - f_1^1(\vec{r}) & f_2^2(\vec{r^*}) - f_2^1(\vec{r})\\
        f_1^2(\vec{r^*}) - f_1^2(\vec{r}) & f_2^2(\vec{r^*}) - f_2^2(\vec{r}) \end{array} \right)$$
with $f_1^1(\vec{r}) = 2^{r_2}(2^{r_1}-1)$, $f_2^1(\vec{r}) = (2^{r_2}-1)$, $f_1^2(\vec{r}) = (2^{r_1}-1)$ and $f_2^2(\vec{r}) = 2^{r_1}(2^{r_2}-1)$

Note that the boundaries between various rate regions are a set of linear equations in $\chi^{-1}$

The maximum value of $<w,r>$ can the be found as
$$ <w,\vec{r}> = \displaystyle\sum_{\pi\in\Pi,\vec{r_i} \in \mathcal{R}_1 \times \mathcal{R}_2} <w,\vec{r_i}> pr(\vec{h} \in S^{\pi}_{\vec{r_i}})$$


For example, in a fixed codebook system with $R_0$ being the rate of the codebook, the optimal scheduler would choose, when the channel state is $\vec{h}$, the maximum of the following :
\begin{enumerate}
\item $0$ (No transmission)
\item $ w R_0 - \kappa_1 \frac{(2^{R_0} - 1)}{h_1^2}$ (Transmission only along channel 1)
\item $(1-w) R_0 - \kappa_2 \frac{(2^{R_0} - 1)}{h_2^2}$ (Transmission along channel 2)
\item $$R_0 - \kappa_1 \frac{(2^{R_0} - 1)}{h_1^2} - \kappa_2 \frac{(2^{R_0} -1)(2^{R_0})}{h_2^2} $$ ( $(R_0,R_0)$ is transmitted with user 2 decoded first )
\item $$R_0 - \kappa_2 \frac{(2^{R_0} - 1)}{h_2^2} - \kappa_1 \frac{(2^{R_0} -1)(2^{R_0})}{h_1^2} $$ ($(R_0,R_0)$ is transmitted with user 1 decoded first)
\end{enumerate}

\begin{figure}[!tbp]
\resizebox{2.6 in}{2.2 in}{\includegraphics*{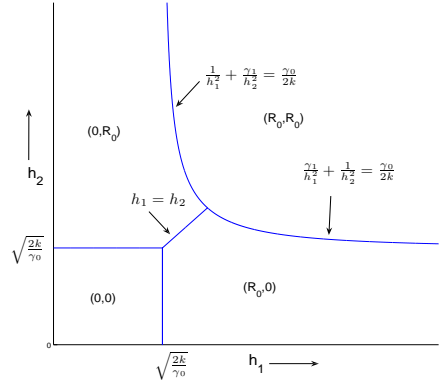}}
\caption{Mapping of channel state to optimal rate vector in MAC network with average power constraints with $w=0.5$, (Lagrangian multipliers assumed to be $\kappa_1 = \kappa_2 = k$) with $\gamma_0 = \frac{R_0}{2^{R_0}-1}$ and $\gamma_1=2^{R_0}-1$}
\label{h_partavgp}
\end{figure}

Figure \ref{h_partavgp} shows the mapping of the optimal rate to the channel state for $w = 0.5$. (Note: for $w = 0.5$, $\kappa_1 = \kappa_2$ because of symmetry). Note that the boundaries between these channel states are linear in $\vec{\chi}^{-1}$ - the Hadamard reciprocal of $\vec{\chi}$.

Note that any increase in $\kappa_i$ has the effect of reducing the corresponding average power consumed $E[P_i(t)]$. In fact, $\kappa_i=0$ corresponds to no power constraint and $\kappa_i \to \infty$ has the effect of $E[P_i(t)] \to 0$. This can be observed in Figure \ref{h_partavgp} where the effect of increasing $k$ will be to increase the size of the region transmitting $(0,0)$ suggesting that the power consumed reduces. This monotonic behavior is exploited in simulations in locating $(\kappa_1^{*},\kappa_2^{*})$ - the Lagrangian multipliers which ensure that the power constraints are satisfied. (Note that in the power constraint the inequality can be replaced by equality). 

The stability region of a fixed codebook MAC networks with $R_0=1$ and average power constraints $\tilde{P_1}$ and $\tilde{P_2}$ satisfying  $\tilde{P_1} + \tilde{P_2} = 2$ are found by performing the above optimization for all $0 \leq w \leq 1$and presented in Figure \ref{resultsfig:avgp}. 


\subsection{BC network}
In the broadcast network, the channel state imposes a decoding order. Therefore, for a given channel state $\vec{\chi}$, the optimal rate allocation is given by 
$$\vec{R}^{*}(\vec{\chi}) = \arg\max_{\vec{r} \in \mathcal{R}_1 \times \mathcal{R}_2} \{<\vec{w},\vec{r}> - \kappa P(\vec{r},\vec{h})\} $$
with $\kappa$ chosen so that the transmitter power constraint is met. 
Since, for a given channel state, the decoding order is fixed, it can be shown that
$$ P(\vec{r},\vec{h}) = \left\{ \begin{array}{ll} \frac{2^{r_2}-1}{h_2^2} + \frac{(2^{r_1}-1) 2^{r_2}}{h_1^2} & \textrm{if } h_1 \geq h_2 \\
\frac{2^{r_1}-1}{h_1^2} + \frac{(2^{r_2}-1) 2^{r_1}}{h_2^2} & \textrm{if } h_2 \geq h_1 \end{array} \right. $$
										
The rate vectors $\vec{r_i} \in \mathcal{R}_1\times\mathcal{R}_2, i=1,2...L$ divide the channel state space into regions $$S_{(r_1^{*},r_2^*)} = \{\vec{\chi}: (r_1^*,r_2^*) = \vec{R}^{*}(\vec{\chi})\}$$. We can write
$$S_{(r_1^{*},r_2^*)} = S^{1}_{(r_1^{*},r_2^*)} \cup S^{2}_{(r_1^*,r_2^*)}$$ where
$$S^{1}_{(r_1^*,r_2^*)}=\{\vec{\chi}: \chi_1 > \chi_2, (r_1^{*},r_2^*) = R^{*}(\vec{\chi})\}$$ and $$S^{2}_{(r_1^*,r_2^*)}=\{\vec{\chi}: \chi_1 \leq \chi_2, (r_1^*,r_2^*) = R^{*}(\vec{\chi})\}$$. It can be shown that, for $k=1,2$
$$S^{k}_{(r_1^*,r_2^*)} = \{\vec{\chi}: A^{(k)}_{\vec{r}} \vec{\chi}^{-1} \leq v, \forall \vec{r} \in 2^{\mathcal{R}_1 \times \mathcal{R}_2} \}$$ where $v=<\vec{w},\vec{r}^{*}-\vec{r}>$ and 
$$ A^{(1)}_{\vec{r}}  = (2^{r_1^*}-2^{r_1},  (2^{r_2^{*}}-1)2^{r_1^{*}}-(2^{r_2}-1) 2^{r_1})$$ 
$$ A^{(2)}_{\vec{r}}  = ((2^{r_1^{*}}-1)2^{r_2^{*}}-(2^{r_1}-1) 2^{r_2}, 2^{{r_2}^*}-2^{r_2})$$ 

The optimal rate allocation for a broadcast network with a fixed codebook of rate $R_0$ chooses, at channel state $\vec{h}$, the maximum of the following:
\begin{enumerate}
\item $$\max\{{0, w R_0 - \kappa \frac{(2^{R_0} - 1)}{h_1^2}, (1-w) R_0 - \kappa \frac{(2^{R_0} - 1)}{h_2^2}}\}$$
\item if $h_1 < h_2$ $$R_0 - \kappa [\frac{(2^{R_0} - 1)}{h_1^2} + \frac{(2^{R_0} -1)(2^{R_0})}{h_2^2}] $$ 
and if $h_1 \geq h_2$, the following quantity
$$R_0 - \kappa [\frac{(2^{R_0} - 1)}{h_2^2} + \frac{(2^{R_0} -1)(2^{R_0})}{h_1^2}] $$ 
\end{enumerate}
The optimal rate allocation partitions the channel state in a manner similar to the MAC ( Figure \ref{h_partavgp}), for $h_1=h_2$ and $w=0.5$.

The stability region of a fixed codebook BC network with $R_0=1$ and $E[P(t)] \leq 2$ is presented in Figure \ref{resultsfig:avgp}.

\section{Duality relationships between stability regions of the MAC and BC networks}
\label{duality}
As discussed in the introduction of this paper, the duality relationship between the information theoretic capacity regions of MAC and BC channels in \cite{goldsmith_duality} provides a motivation to explore similar results with our system model. In the discussion that follows, a Gaussian MAC network is defined to be the \textit{dual} of a BC network with identical number of users if
\begin{enumerate}
\item The power constraints of the MAC and BC networks have the same form  i.e either both have average power constraints or peak power constraints
\item The maximum transmit power (average or peak, as the case may be) of the broadcast network is equal to the sum of the maximum transmit powers of all users of the MAC network (For example, the dual MAC network of a two user broadcast network with power constraints $P(t) \leq \tilde{P}$ has power constraints $P_1(t) \leq \tilde{P_1}$ and $P_2 \leq \tilde{P_2}$ such that $\tilde{P_1}+\tilde{P_2}=\tilde{P}$ )
\item Their channel states have identical statistics and the noise variance at all the receivers in the BC network are identical to the noise variance at the receiver in the MAC.
\item The set of rates of the codebooks available for all the MAC users are identical to each other and are identical to the set of rates at the transmitter in the broadcast channel.
\end{enumerate}

We reproduce MAC-BC duality results from \cite{goldsmith_duality} for completeness.

\begin{theorem}
\label{phyduality}
[Jindal et. al.] For a fixed channel state, the information theoretic capacity region of a Gaussian Broadcast channel with power constraint $\bar{P}$ is equal to the union of the capacity regions of the dual multiple access channel with power constraints $(\tilde{P_1},\tilde{P_2},..\tilde{P_K})$ such that $\displaystyle\sum_{j=1}^{K} \tilde{P_j} = \tilde{P}$. Furthermore, every point on boundary of the capacity region of the broadcast channel is the \textit{corner point} of a pentagon representing the capacity region of some dual MAC channel.
\end{theorem}


The fact that the boundary of the capacity region of the BC channel is achieved by the \textit{corner points} of the MAC capacity regions is important in the duality between stability regions (below) since in our system model, only the corner points of the pentagon representing the MAC capacity regions are achievable. 

The relationship between the \textit{stability regions} of MAC and BC networks are characterized in Theorems \ref{subset_duality} and \ref{avg_duality} below.

\begin{theorem}
\label{subset_duality}
The union of stability regions of all MAC networks that are dual to a particular BC network with peak power constraint is \emph{strictly contained in} the stability region of that dual BC network.
\end{theorem}
\begin{proof}
Refer Appendix \ref{proof:peak_duality} for an explanation based on Theorem \ref{phyduality}.
\end{proof}
However, a duality relationship in the form of equality holds between stability regions of peak power constrained MAC and BC networks of a special type - MAC networks with \textit{centralized power} i.e the power can be shared between the two transmitters of the MAC network and the power constraint is expressed as $P_1(t)+P_2(t) \leq \tilde{P}$. At a given channel state $(h_1,h_2)$, the physical layer capacity duality result implies that, for every rate vector transmitted by the broadcast channel, there is at least one multiple access channel whose sum power is equal to $\tilde{P}$ that can achieve that rate. Therefore at time slot $t$, powers $P_1(t)$ and $P_2(t)$ in the centralized-power MAC network can be picked so as to emulate the appropriate `distributed power' multiple access channel to achieve the rate vector in the corresponding dual broadcast channel. In other words, the MAC channel with centralized power can switch between different dual MAC channels depending on the channel state, to achieve the rate vector transmitted in the broadcast network.

For MAC and BC networks with average power constraints, a duality result similar to the one presented in \cite{goldsmith_duality} holds.

\begin{theorem}
\label{avg_duality}
The stability region of a broadcast network with an average transmit power constraint \emph{is equal to} the union of the stability regions of all the its dual MAC networks.
\end{theorem}

\begin{figure}[!tbp]
\resizebox{2.8 in}{2.4 in}{\includegraphics*{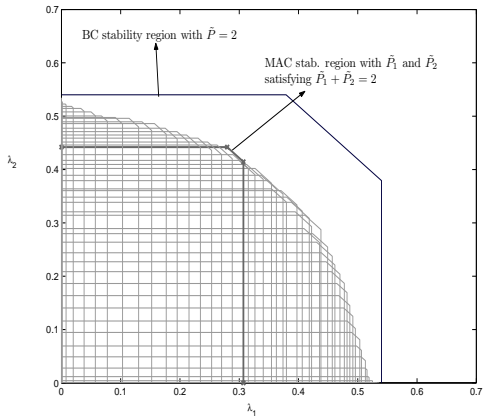}}
\caption{Stability regions of the BC and several dual MAC networks with peak power constraints}
\label{resultsfig:peakp}
\end{figure}

\begin{figure}[!tbp]
\resizebox{2.7 in}{2.1 in}{\includegraphics*{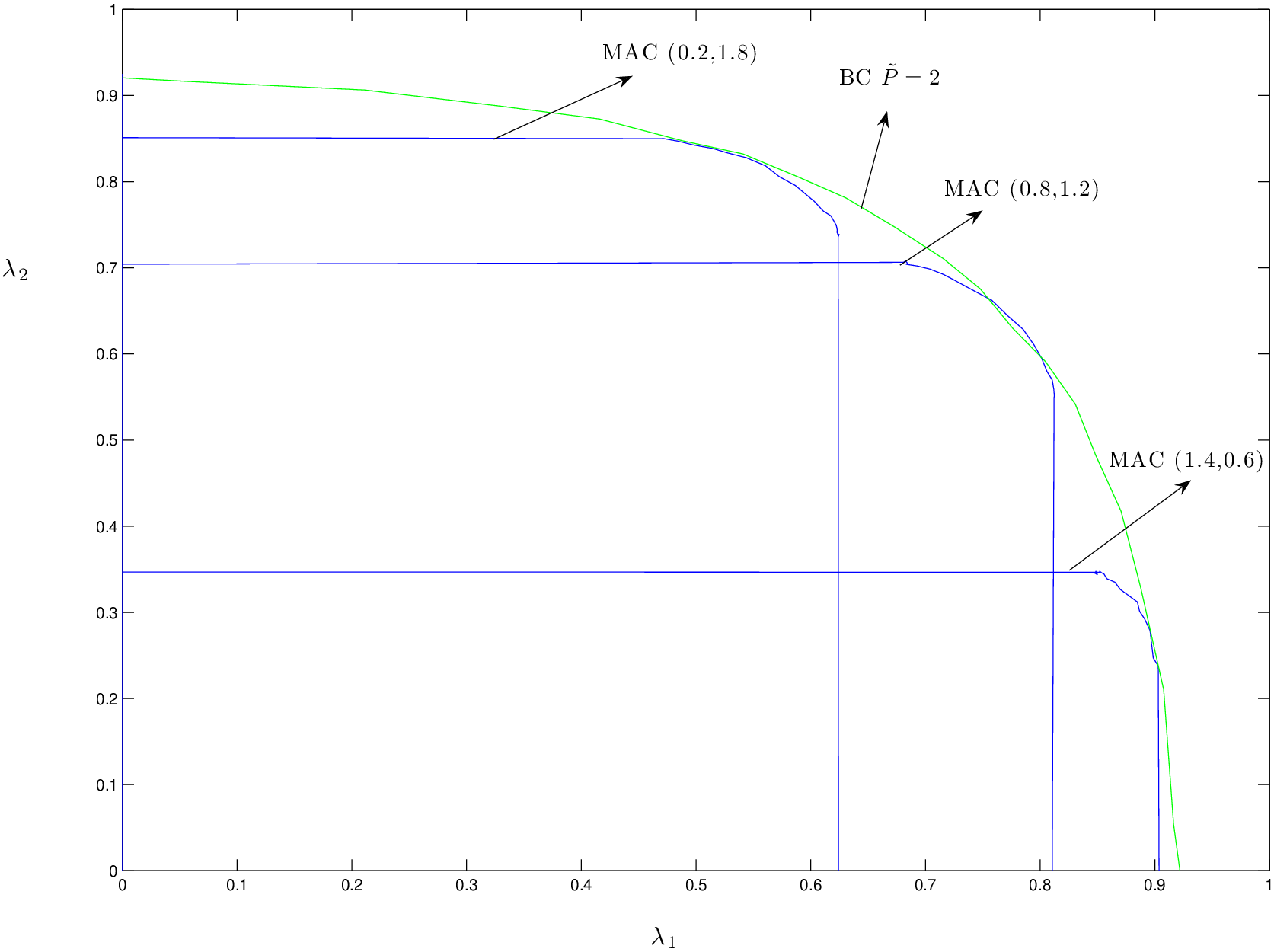}}
\caption{Stability regions of the BC and several dual MAC networks with average power constraints}
\label{resultsfig:avgp}
\end{figure}

Interestingly, the duality result holds . Proofs for Theorems \ref{subset_duality} and \ref{avg_duality} are provided in the Appendices \ref{proof:peak_duality} and \ref{proof:avg_duality} .

The plots of stability regions of BC and dual MAC networks with peak and average power constraints (Figures \ref{resultsfig:peakp} and \ref{resultsfig:avgp}) confirm the results established in Theorem \ref{subset_duality} and \ref{avg_duality}.

From the plot, it can be observed that in the peak power constraint case, the convex hull of all the MAC pentagons, representing time division multiplexing between the different multiple access channels, is a subset of the broadcast pentagon. Note that the convex hull of the MAC stability regions is different from the stability region of the MAC networks with centralized power; the former represents the stability region achieved by choosing a MAC channel among the various dual MAC channels, \textit{randomly}. Whereas the latter stability region that can be achieved if a scheduler can choose a dual MAC channel, based on the channel conditions. 

\section{Optimal Codebooks}
\label{optimal_codebooks}
There is an interesting extension associated with the stability region problem in \emph{multi-rate} MAC and BC networks.  Consider a general $N$ user MAC network. Given the number of rates allowed at each user i.e given $|\mathcal{R}_1|,|\mathcal{R}_2|...|\mathcal{R}_N|$, how do we choose rates $\mathcal{R}_i, i=1,...N$ so as to maximize the sum-rate point in the stability region ? The problem clearly has important applications in the design of communication systems.

To simplify the problem, consider a MAC network with identical peak power constraints at all the users. Also, assume that all users have a codebook of rate $R_0$ i.e $|\mathcal{R}_i| = 1, i=1,2...N$. Given $N$ and $R_0$, the stability region of the network can be found using techniques presented in section \ref{stability:peakp}. Let $s(R_0,N)$ indicate the maximum stable sum-rate i.e $$ s(R_0,N) = \max \displaystyle\sum_{i=1}^{N} \bar{R_i}$$ under the constraint that $(\bar{R_1},...\bar{R_N})$ lies in the stability region. 
The optimal codebook design problem can be formulated as follows. 
$$R_0^{*}(N) = \arg\max_{R_0} s(R_0,N)$$
\begin{figure}[!tbp]
\resizebox{2.6 in}{2.2 in}{\includegraphics*{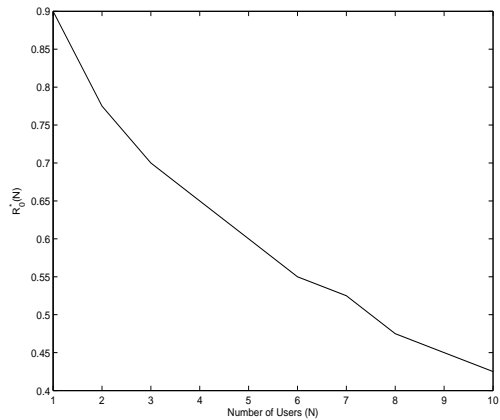}}
\caption{Optimal codebook rate $R_0^*(N)$ in a fixed codebook MAC network versus number of users $N$}
\label{optcodebook_mac}
\end{figure}

\begin{figure}[!tbp]
\resizebox{2.6 in}{2.2 in}{\includegraphics*{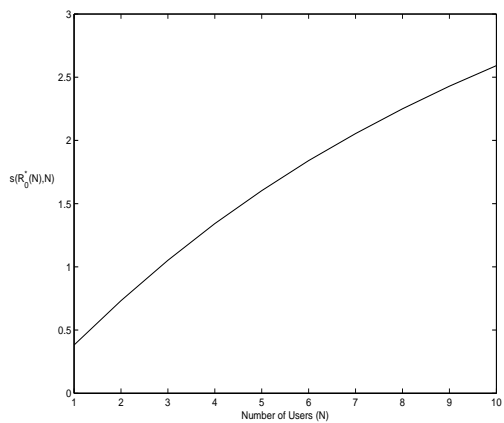}}
\caption{Optimal value of sum-rate $s(R_0^*(N),N)$ in a MAC network versus number of users $N$}
\label{sumrate_mac}
\end{figure}

\begin{figure}[!tbp]
\resizebox{2.6 in}{2.2 in}{\includegraphics*{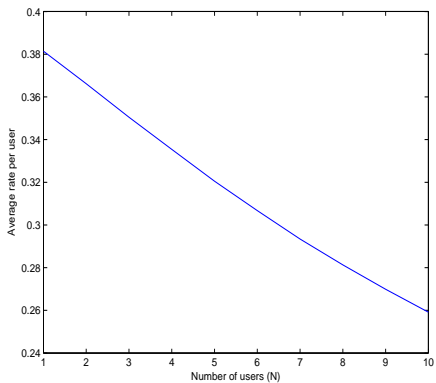}}
\caption{Average rate per user versus the number of users with the optimal codebook in a MAC network }
\label{sumratefairness_mac}
\end{figure}

Plots of $R_0^{*}(N)$ (found by simulations iterating over different $R_0$) and the corresponding sum-rate $s(R_0^{*}(N),N)$ are provided in Figures \ref{sumrate_mac} and \ref{optcodebook_mac} respectively. Rayleigh fading of unit variance, and unit power at all the transmitters are used in the simulation. Note that Figure \ref{sumrate_mac} plots, versus $N$, the highest possible stable sum-rate in fixed codebook N-user MAC network. The increase in $s(R_0^{*}(N),N)$ can be trivially explained - a sum rate of $s(R_0,N-1)$ can be achieved in a N user system, using a scheduler that simply ignores the $N$th user. The behavior of $R_0^{*}(N)$ with increasing $N$ is not known for a general channel distribution, though the plot in \ref{optcodebook_mac} indicates that it decreases with increasing $N$ in Rayleigh fading channels. Also, the plot in Figure \ref{sumratefairness_mac} suggests that the rate seen by each user decreases with increasing $N$ (This can also be observed simply by the fact that the rate of increase of $s(R_0^*(N),N)$ in Figure \ref{sumrate_mac} is decreasing). This suggests a trade-off between the the throughput of the system and the average rate seen by each user.

\section{Conclusions and Future Work}
\label{conclusion}
A framework combining both information theoretic and networking points of view of MAC and BC networks has been presented and their stability regions have been characterized.  The assumptions on encoding and decoding in the system model implied that the information theoretic capacity region did not present the complete picture and this led to some interesting observations. For example, we found that for networks with peak power constraint, the optimal scheduling policy may transmit at a power less than the maximum possible power in peak-power constrained MAC networks. 

Stability of MAC and BC networks with both average and peak power constraints were found. The stability regions of MAC and BC networks with average power constraints satisfy the fundamental duality property that relates the information theoretic capacity regions of these networks.  Interestingly, the duality property holds in spite of restricting users to finite rate-sets and using a suboptimal decoding strategy.  In the peak power constraint case, the union of MAC networks is found to be a subset of the corresponding dual BC networks. 

This work opens up some interesting areas of future research. Characterization of stability regions of more complicated networks such as the interference, X and Z channels, and relay networks can potentially provide interesting insight. The information theoretic capacity regions of the networks mentioned are not yet known; however, decoding scheme can be restricted successive interference cancellation and their stability regions can be determined. Thus stability regions could be useful metrics for these communication networks. Their stability regions could possibly share some fundamental properties with their capacity regions like the duality property in MAC and BC networks. Therefore, a study of their stability regions may potentially provide useful information about the nature of their capacity regions.
Another area of future work is the design of optimal codebooks for these networks discussed in Section \ref{optimal_codebooks}. While Section \ref{optimal_codebooks} discussed the fixed codebook version of the problem, an interesting issue is the loss in performance by restricting users to a finite number of codebooks in comparison to the information theoretic capacity region. In other words, a characterization of the number codebooks that should be used so that the sum-rate achieved is `close' to the maximum sum-rate in the information theoretic capacity region is an important related problem in the design of communication networks.




\nocite{dtse:mac}
\nocite{ephremides:lyapunov}
\bibliographystyle{unsrt}
\bibliography{refs}

\begin{thebibliography}{10}

\bibitem{thomas_cover}
Thomas~M. Cover and Joy~A. Thomas.
\newblock {\em Elements of information theory}.
\newblock Wiley-Interscience, New York, NY, USA, 1991.

\bibitem{gallagher:networks}
D.~Bertsekas and R.~Gallagher.
\newblock {\em {Data Networks}}.
\newblock Prentice Hall, Englewood Cliffs, NJ, 2nd edition, 1991.

\bibitem{mlwdf}
Matthew Andrews, Krishnan Kumaran, Kavita Ramanan, Alexander Stolyar, Phil
  Whiting, and Rajiv Vijayakumar.
\newblock Providing quality of service over a shared wireless link.
\newblock {\em IEEE Communications Magazine}, pages 150--154, February 2001.

\bibitem{mneely_thesis}
M.~Neely, E.~Modiano, and C.~Rohrs.
\newblock Dynamic power allocation and routing for time varying wireless
  networks, 2003.

\bibitem{mneely:avgp}
Michael~J. Neely.
\newblock Energy optimal control for time varying wireless networks.
\newblock In {\em INFOCOM}, pages 572--583, 2005.

\bibitem{goldsmith_duality}
A~Goldsmith, Sriram Vishwanath, and N~Jindal.
\newblock On the duality of gaussian multiple-access and broadcast channels.
\newblock {\em IEEE Transactions on Information Theory}, 50:768--783, 2004.

\bibitem{3gpp}
3gpp2.
\newblock cdma2000 evaluation methodology - version 1.0, 3gpp2 cr1002-0 v1.0,
  December 2004.
\newblock http://www.3gpp2org/Public\_html/specs.

\bibitem{ieee:edge}
A.~Furuskar, S~Mazur, F~Muller, and H~Olofsson.
\newblock Edge: enhanced data rates for gsm and tdma/136 evolution.
\newblock {\em IEEE Personal Communications}, 6:56--66, June 1999.

\bibitem{cross_survey1}
Sanjay Shakkotai, Theodore~S Rappaport, and Peter~C Karlsson.
\newblock Cross-layer design for wireless networks.
\newblock {\em IEEE Communications Magazine}, pages 74--80, October 2003.

\bibitem{cross_survey2}
Vineet Srivatsava and Mehul Motani.
\newblock Cross-layer design: a survey and a road ahead.
\newblock {\em IEEE Communications Magazine}, pages 112--119, February 2005.

\bibitem{eyeh:routing_relay}
Edmund Yeh and Randall Berry.
\newblock Throughput optimal control of cooperative relay networks.
\newblock In {\em International Symposium on Information Theory}, September
  2005.

\bibitem{mneely:book}
Leonidas Georgiadis, Mike Neely, and Leandros Tassiulas.
\newblock {\em Resource Allocation and Cross Layer Control in Wireless Networks
  (Foundations and Trends in Networking, V. 1, No. 1)}.
\newblock Now Publishers Inc, 2006.

\bibitem{ephremides:randommac}
Jie Luo and Anthony Ephremides.
\newblock On the throughput, capacity, and stability regions of random multiple
  access.
\newblock {\em IEEE/ACM Trans. Netw.}, 14(SI):2593--2607, 2006.

\bibitem{ltong:aloha}
V.~Naware, G.~Mergen, and L.~Tong.
\newblock Stability and delay of finite user slotted aloha with multipacket
  reception.
\newblock {\em IEEE Transcations on Information Theory}, 51:2636-- 2656, July
  2005.

\bibitem{ltong:capstab}
G{\"o}khan Mergen and Lang Tong.
\newblock Stability and capacity of regular wireless networks.
\newblock {\em IEEE Transactions on Information Theory}, 51(6):1938--1953,
  2005.

\bibitem{telatar_qthmac}
I~E Telatar and R~G Gallager.
\newblock Combining queueing theory with information theory for multiaccess.
\newblock {\em {IEEE} {J}ournal on {S}elected {A}reas in {C}ommunications},
  13(6):963--969, 1995.

\bibitem{ephremides_infotnetworksunion}
Anthony Ephremides and Bruce~E. Hajek.
\newblock Information theory and communication networks: An unconsummated
  union.
\newblock {\em IEEE Transactions on Information Theory}, 44(6):2416--2434,
  1998.

\bibitem{eyeh_macnetcap}
Edmund Yeh and AS~Cohen.
\newblock Throughput optimal power and rate control for queued multiaccess and
  broadband communications.
\newblock {\em Proceedings of International Symposium on Information Theory},
  2004.

\bibitem{neely:book}
Leonidas Georgiadis, Michael~J. Neely, and Leandros Tassiulas.
\newblock Resource allocation and cross-layer control in wireless networks.
\newblock {\em Found. Trends Netw.}, 1(1):1--144, 2006.

\bibitem{syed_cdmamultirate}
S.~Jafar and A.~Goldsmith.
\newblock Optimal rate and power adaptation for multirate cdma.
\newblock In {\em Proc. IEEE Veh. Tech. Conf. (VTC), Boston}, 2000.

\bibitem{dtse:mac}
D.N.C. Tse and S.V Hanly.
\newblock Multiaccess fading channels. i. polymatroid structure, optimal
  resource allocation and throughput capacities.
\newblock {\em IEEE Transactions on Information Theory}, Vol.44(7), 1998.

\bibitem{ephremides:lyapunov}
A~Ephremides and L.~Tassiulas.
\newblock Stability properties of constrained queueing systems and scheduling
  policies for maximum throughput in multihop radio networks.
\newblock In {\em Proceedings, 29th IEEE Conference on Decision and Control},
  volume~4, December 1990.

\end{thebibliography}
\appendices
\section{Proof of Lemma 1}
\begin{proof}
\label{proof:avgp_lemma}
The fact that a channel state can be uniquely mapped to a single rate in the optimal policy in networks with average power constraints can be proved for general N-user MAC and BC networks with multiple rate-sets but such a proof is complicated. The idea is conveyed in the proof of the result for a fixed codebook point-to-point link. 

Consider a point-to-point link with average power constrained by $E[P(t)] \leq \tilde{P}$. The link has a fixed codebook of rate $R_0$. The channel is Gaussian with gain $h(t)$ which takes values from a continuous state space. 
At a given state, $h$, the only parameter that affects the average rate and power is the probability of transmission at this state $p(h)$. Note that the effect of any algorithm, in terms of average rate and power consumed can be achieved by an equivalent randomized algorithm with the appropriate `weight' function $p(h)$. The problem of finding the stability region of this network is simply equivalent to
$$\max_p(h)\int R_0 p(h)f(h)dh$$
s.t $\int P(h) p(h)f(h)dh \leq P$\\ where
$P(h)$ is the power required to transmit a codeword at state $h$. Note that $P(h)$ is monotonically decreasing in $h$.

We intend to prove that in the optimal policy a given channel state is mapped to a single rate vector in $\{R_0,0\}$ or equivalently, that $p(h) \in \{0,1\}, \forall h$

Let the optimal rate allocation policy be $p^*(h)$. Contrary to our claim, let $p^{*}(h)$ take values in $(0,1)$ for certain $h_0 \in \mathbb{R}$.
Now, consider an interval $(x,y)$ along the positive axis on the real line such that $p(h) \notin \{0,1\}, \forall h \in (x,y)$ and $\int_x^y f(h) > 0$ . Note that if we cannot find such an interval, then setting all values of $p'(h) = \lfloor p(h) \rfloor$  is also optimal and satisfies the given property.
We now find an alternate scheduling policy $p_1(h)$ which performs better than $p^{*}(h)$ as a contradiction. \\We know that since $p^{*}(h)<1,\forall h\in(x,y)$
$$\int_{x}^{y} p^*(h) P(h) f(h) dh < \int_{x}^{y} P(h) f(h) dh $$
\begin{equation} \label{equation:m}\therefore  \exists z \in (x,y) \mbox{ s.t } \int_x^y p^{*}(h) P(h) f(h) = \int_z^{y} P(h) f(h) dh\end{equation}
 $$\Rightarrow \int_x^{z} P(h) p^{*}(h) f(h) dh = \int_z^{y} P(h) (1-p^{*}(h)) f(h) dh $$
Since $P(h)$ is decreasing in $h$, we have
$$ P(z) \int_x^{z} p^{*}(h) f(h) dh < \int_z^{y} P(h) (1-p^{*}(h)) f(h) dh$$ and
$$  \int_z^{y} P(h)(1-p^{*}(h)) f(h) dh <  P(z) \int_z^{y} (1-p^{*}(h)) f(h) dh$$ 
Combining the above two inequalities, we have
 $$\int_x^{z} p^{*}(h) f(h) dh < \int_z^{y} (1-p^{*}(h)) f(h) dh$$
\begin{equation} \label{equation:m1}\Rightarrow \int_x^{y} p^{*}(h) f(h) dh < \int_z^{y} f(h) dh \end{equation}
Now, we construct a new scheduling policy as follows
$$ p_1(h) = \left\{ \begin{array}{ll} p^{*}(h) & \mbox{if } h \notin(x,y)\\
0 & \mbox{if } h \in[x,z)\\
1 & \mbox{if } h \in[z,y] \end{array} \right.$$
Equation \ref{equation:m} implies that the power consumed with $p_1(h)$ is the same as the power consumed with $p^{*}(h)$. Equation \ref{equation:m1} implies that the average rate achieved with $p_1(h)$ is greater than that achieved with $p^{*}(h)$, contradicting the assumption that $p^{*}(h)$ is optimal. The result is hence proved. 
\end{proof}

\section{Proof of Theorem \ref{subset_duality}}
\label{proof:peak_duality}
\begin{proof}
We observe the channel state space $(h_1(t), h_2(t))$ has identical statistics for the broadcast and the dual MAC channel.  Therefore, if we show that, for a given channel state, all rate vectors that can be supported by the MAC channel can be supported by the dual broadcast channel, then the stability region of the multiple access network is a subset of the network capacity region of the broadcast network (since the BC network can follow use the appropriate scheduling strategy and achieve the corresponding rate vector) . This is precisely the result stated by the physical layer duality result (Theorem \ref{duality}), and hence the result follows.

In fact, the union of MAC stability regions is a proper subset of the BC stability region. To see this, 
we perform the following thought experiment. Considering a two-user fixed codebook broadcast channel with a peak power constraint $\tilde{P}$ whose channel gains $(h_1,h_2)$ can, take two states, $ S_1 = (h_{11},h_{21})$ and $ S_2 = (h_{12},h_{22})$ with equal probability (of $\frac{1}{2}$) such that $$h_{12} = h_{21} < h_{11} = h_{22}$$ and  
$$\log(1+h_{11}^2 \tilde{P}) = R_0$$
where $R_0$ is the rate of the codebook in the BC channel. Therefore, we have :
$$\log(1+h_{12}^2 \tilde{P}) < R_0$$
In other words, there two possible channel states. In each channel state, one of the two channels is \textit{ON} and the other is \textit{OFF}. The corresponding set of supported rates are $\{(R_0,0)\}$ and $\{(0,R_0)\}$ (assuming additive white Gaussian noise of unit variance at each receiver). Using the result of equation \ref{eqn_stab_peakp}, the stability region of this BC network is a square in the Cartesian plane (see Figure \ref{figure:duality_subset}) formed by the points $(0,0)$, $({R_0},0)$, $(0,{R_0})$, $({R_0},{R_0})$.  

Among the set of all dual multiple access networks with power constraint $\tilde{P_1}$ and $\tilde{P_2}$ satisfying $ \tilde{P_1} + \tilde{P_2} = \tilde{P}$, transmission is possible only if $\tilde{P_1}=\tilde{P}$ or $\tilde{P_2} =\tilde{P}$. In other words, if $\tilde{P_1} < \tilde{P}$ and $\tilde{P_2} < \tilde{P}$, no transmission is possible. In the MAC network with power constraints $(\tilde{P},0)$, the sets of supported rates are $\{(R_0,0)\}$ for channel state \textit{ON} and $(0,0)$ for channel state \textit{OFF}. The stability region is therefore a line joining $(0,0)$ and $(\frac{R_0}{2},0)$. The union of stability regions of dual MAC channels is thus, the union of two line segments along the axes forming the adjacent sides of the square representing the BC stability region. Clearly, this is a proper subset of the BC stability region. This proper subset relationship is confirmed by results in Figure \ref{resultsfig:peakp}.

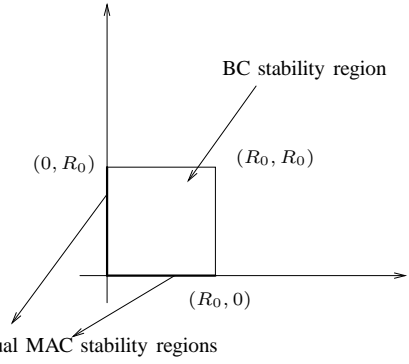
\begin{figure}[!tbp]
{\setlength{\unitlength}{0.00046667in}
\begingroup\makeatletter\ifx\SetFigFont\undefined%
\gdef\SetFigFont#1#2#3#4#5{%
  \reset@font\fontsize{#1}{#2pt}%
  \fontfamily{#3}\fontseries{#4}\fontshape{#5}%
  \selectfont}%
\fi\endgroup%
{\renewcommand{\dashlinestretch}{30}
\begin{picture}(5637,3985)(0,-10)
\path(2355.000,3838.000)(2325.000,3958.000)(2295.000,3838.000)
\path(2325,3958)(2325,658)
\path(2025,958)(5625,958)
\path(5505.000,928.000)(5625.000,958.000)(5505.000,988.000)
\path(2325,2158)(3525,2158)(3525,958)
\drawline(3075,958)(3075,958)
\thicklines
\path(2325,958)(3525,958)
\path(2325,958)(2325,2158)
\thinlines
\path(3975,3058)(3225,2008)
\path(3270.337,2123.085)(3225.000,2008.000)(3319.161,2088.211)
\path(2325,1858)(1275,433)
\path(1322.032,547.403)(1275.000,433.000)(1370.336,511.811)
\path(3075,958)(1950,283)
\path(2037.464,370.464)(1950.000,283.000)(2068.334,319.015)
\put(3750,2158){\makebox(0,0)[lb]{{\SetFigFont{7}{8.4}{\familydefault}{\mddefault}{\updefault}$(R_0,R_0)$}}}
\put(3225,583){\makebox(0,0)[lb]{{\SetFigFont{7}{8.4}{\familydefault}{\mddefault}{\updefault}$(R_0,0)$}}}
\put(1500,2083){\makebox(0,0)[lb]{{\SetFigFont{7}{8.4}{\familydefault}{\mddefault}{\updefault}$(0,R_0)$}}}
\put(3600,3133){\makebox(0,0)[lb]{{\SetFigFont{8}{9.6}{\rmdefault}{\mddefault}{\updefault}BC stability region}}}
\put(0,58){\makebox(0,0)[lb]{{\SetFigFont{8}{9.6}{\rmdefault}{\mddefault}{\updefault}Union of dual MAC stability regions}}}
\end{picture}
}}
\caption{Stability regions of BC and dual MAC networks with `ON-OFF' channels}
\label{figure:duality_subset}
\end{figure}
\end{proof}

\section{ Proof for Theorem \ref{avg_duality}} 
\label{proof:avg_duality}
\begin{proof}
In a BC network with average power constraints, the optimal policy schedules a rate vector $\vec{\mu}(\vec{h})$ when the channel state is $\vec{h}$.  Let the power transmitted at channel state $\vec{h}$ be $P(\vec{h})$. We provide a MAC network and a scheduling policy for this network that achieves the following:
\begin{enumerate}
\item The average rate vector as a result of the scheduling policy is equal to the average rate vector of the BC network
\item The average power expended by the users in the MAC network satisfy the duality property i.e, the sum of the average powers consumed is equal to the average power expended in the BC network.  
\end{enumerate}
Existence of such as scheduling policy ensures that the stability region of the broadcast network is a subset of the union of stability regions of dual MAC networks.

We construct the scheduling policy as follows. Since $\vec{\mu}$ lies in the capacity region of the broadcast network (for $\vec{h}$),  we know from Theorem \ref{phyduality} (Section \ref{duality}) that there exists a  power allocation $(P_1,P_2)$ in the MAC network we have considered, such that $P_1+P_2 = P(\vec{h})$ and $\vec{\mu}(h)$ can be transmitted successfully. Note the capacity region of the broadcast channel is achieved by the corner points of the pentagons representing physical layer capacity regions of the dual MAC channels. Therefore, in spite of disallowing rate-splitting, we can find $(P_1(h),P_2(h))$ in the MAC channel so that $\vec{\mu}(h)$ can be transmitted. Also, for a given $\vec{h}$, $P_1$ and $P_2$ are purely functions of the rate vector $\vec{\mu}(h)$ transmitted by the broadcast channel. Let $P_1(h) = f(\vec{h},\vec{\mu}(h))$ and $P_2(h) = g(\vec{h},\vec{\mu}(h))$. Now the scheduler in the MAC networks transmits $\vec{\mu}(h)$ with the two users using powers $f(\vec{h},\vec{\mu}(h)))$ and $g(\vec{h},\vec{\mu}(h))$. Clearly this MAC network can stabilize any arrival rate stabilized by the broadcast network. Also, since $E[P_1(h)] + E[P_2(h)] = E[P_1(h)+P_2(h)] = E[P(h)]$, and therefore $E(g(\vec{h},\vec{\mu}(h))) + E(f(\vec{h},\vec{\mu}(h))) = \tilde{P}$, the power constraints of the MAC network satisfy the duality constraint. Therefore, the broadcast network stability region is a subset of the union of stability regions dual multiple access channels with average power constraints. An analogous argument can be used to show that the union of stability regions of dual MAC networks is a subset of the stability region of the broadcast network. Thus the stability region of a broadcast network with average power constraint is equal to the union of stability regions of dual MAC networks.
\end{proof}
\end{document}